\documentclass[12pt]{article}
\usepackage{subfigure}
\usepackage{url}
\usepackage{color}
\usepackage{xcolor}
\usepackage{graphicx}
\usepackage{amsmath}%
\usepackage{subcaption,lipsum}
\usepackage{amssymb}
\newcounter{bla}

\newcommand{\ddt}[1]{\frac{d{#1}}{dt}}
\graphicspath{{./figs/}}

\usepackage[a4paper, left=2.5cm, right=2.5cm, top=3cm, bottom=3cm]{geometry} 
\usepackage{authblk}

\newcommand{\REDLIB}{RePlaChem}

\title{\REDLIB: A dimensionality reduction library for plasma chemical mechanisms.}

\author[a]{Z. Nikolaou \footnote{Corresponding author: nikolaou.zacharias@ucy.ac.cy}}
\author[b]{E. Morais}
\author[b,c]{S. Van Rompaey}
\author[a]{C. Anastassiou}
\author[b]{A. Bogaerts}
\author[a,d]{V. Vavourakis}

\affil[a]{In Silico Modelling Group, Department of Mechanical \& Manufacturing Engineering, University of Cyprus, Cyprus.}
\affil[b]{PLASMANT, and Center of Excellence PLASMA, Department of Chemistry, University of Antwerp, Antwerp, Belgium.}
\affil[c]{Research Unit Plasma Technology (RUPT), Department of Applied Physics, Sint-Pietersnieuwstraat 41, Ghent University, Ghent, Belgium}
\affil[d]{Department of Medical Physics \& Biomedical Engineering, University College London, London, United Kingdom}

\begin{document}

\maketitle

\hrule

\begin{abstract}
In this work, we present \REDLIB, a software library for reducing detailed large-scale plasma chemical mechanisms to smaller skeletal ones. The library parses a plasma chemical mechanism in the well-established format compatible with the software ZDPlasKin, runs the reduction algorithm, and generates automatically a skeletal chemical mechanism. This feature allows the seamless implementation of the skeletal chemistry in well-known solvers including ZDPlasKin. In turn, the reduction of the chemistry accelerates otherwise computationally expensive numerical simulations. Furthermore, \REDLIB\ can be used as an analysis tool to shed light into the underlying reaction physics by identifying the dominant species and associated reactions. In order to validate and demonstrate its capabilities, \REDLIB\ is used to substantially reduce a large-scale methane-hydrogen plasma chemical mechanism with 77 species and 4404 reactions at various orders with a relatively small loss of accuracy. \\ 
\textit{GitHub: \url{https://github.com/znikolaou/RePlaChem.git}}

\end{abstract}

\newpage

\section{Introduction}

Plasmas occur naturally on earth (lightning, Aurora Borealis), throughout the universe (nebula, solar corona, ionic-wind), and are also generated artificially either by heating or by applying strong electromagnetic fields to a neutral gas. In industry, plasmas have numerous and diverse applications. Hall-effect thrusters for instance in satellites, utilise plasma jets \cite{2008_wiley_goebel}. Plasma torches are used for welding/cutting, and semiconductors in electronic devices are treated (etching, deposition) using plasma \cite{1993_applPhysLet_ventzek}. Plasmas also have some remarkable medical applications. Non-thermal plasma medical devices are currently being used for wound-healing \cite{2016_clinicalPlasmaMed_sander}, and are also considered for treating cancer \cite{2013_physPlasmas_keidar}. Another major application of artificial plasmas is in the chemical industry. Ozone, used for the removal of pollutants and micro-organisms in gas and liquid mixtures, is produced at scale using electric discharge plasma \cite{1988_procTech_kogel}. Plasma has also shown great potential for the electrification of traditional thermocatalytic processes, such as reforming methane and carbon dioxide, methane coupling and pyrolysis, nitrogen fixation for sustainable fertiliser production, ammonia cracking to name but a few applications \cite{2017_chemSoc_snoeckx,2018_acs_bogaerts,2020_heijkers_jPhysChem,2021_joule_winter,FULCHERI2024100973,FEDIRCHYK2024155946,2025_natureChem_bogaerts}. For an in-depth review on plasma applications for sustainable gas conversion, we refer the reader to the latest plasma roadmap \cite{2022_jPhysD_adamovich}. 

Inevitably, plasma chemistry plays an important role \cite{2008_plasmaChem_fridman}, and coupled multi-physics numerical simulations aim to provide accurate predictions of both the flow dynamics and the species concentrations in time/space (or the more commonly used number densities). To achieve that, a range of space and time scales must be resolved governing plasma transport, fluid transport, volume and surface reactions etc. \cite{2009_jPhysD_kushner, 2018_psst_trelles}. Large-scale detailed chemical mechanisms may contain tens to hundreds of species as well as hundreds to thousands of reactions  \cite{gaens_jPhysD_2013,2016_jPhysD_leitz,2018_physChem_schroter}. Resolving a multitude of species with vastly differing time-scales is a challenging computational problem, and in most simulations time-steps of the order of ns are typically used which increases the computational workload substantially. This makes detailed-chemistry plasma simulations quite expensive and perhaps even prohibitive in 3D \cite{2018_psst_trelles}-in fact most simulations to date are limited to 2D and using smaller-scale chemical mechanisms.  

The same issue has long plagued the combustion modelling and simulation community. To address this, numerous and diverse strategies were developed throughout the years. Classic methods include Sensitivity Analysis \cite{1989_intJChemKin_turanyi}, reaction-lumping \cite{2008_ctm_pepiot}, Quasi-Steady-State-Assumptions \cite{2006_jPhysChem_lu,2013_cnf_nikolaou,2014_cnf_nikolaou}, Directed Relation Graphs (DRG) \cite{2005_procCombustInst_lu} and Directed Relation Graphs with Error Propagation (DRGEP) \cite{2008_cnf_pepiot}, Intrinsic Low Dimensional Manifolds (ILDM) \cite{1992_cnf_maas}, Computational Singular Perturbation \cite{1994_intJChemKin_lam}, and Principal Component Analysis \cite{2009_pca_cnf_sutherland} to name but a few. In more recent works, neural networks have also been employed \cite{2020_anns_cnf_wan} while a further powerful strategy is to use optimisation as well \cite{furst_compPhysComm_2021}. Another approach, is to directly accelerate the integration of the species concentrations by off-loading to Graphics Processing Units (GPUs) and/or generally exploiting the system architecture \cite{niemeyer_compPhysComm_2017,curts_compPhysComm_2022,rao_compPhysComm_2024,danciu_compPhysComm_2025}. In-Situ Adaptive Tabulation \cite{2010_ctm_pope_isat} is also an alternative acceleration approach.

In the plasma physics community, there are many tools for simulating plasma flows \cite{dechant_compPhysComm_2023,verma_compPhysComm_2021}, and for plasma chemical kinetics \cite{zdplaskin,pinhao_compPhysComm_2001,bolsigSolver}. However, the literature on reduction is much more limited in comparison to combustion. A Reaction Pathway Analysis (RPA) method, originally proposed for analysing reaction pathways in atmospheric chemistry \cite{2004_jAtmosphChem_lehman}, was first implemented in \cite{markosyan_compPhysComm_2014} using the software tool PumpKin. RPA requires a number of inputs from the user: the reaction-rates, the species number-density profiles in time, an estimate for the lifetime of the target species, and a threshold rate value for eliminating un-important reactions. PumpKin additionally requires the user to provide the species stoichiometric coefficients-a rather daunting task for large-scale complex chemical mechanisms. Despite the works in \cite{2004_jAtmosphChem_lehman,markosyan_compPhysComm_2014} the method's capabilities as a reduction rather as an analysis tool were investigated much later in \cite{2022_jPhysD_mousavi}. After proposing some modifications to overcome some of the method's original short-comings \cite{2022_jPhysD_mousavi}, RPA was used to reduce two sets of relatively small-scale chemical mechanisms (622 reactions with 56 species, and 306 reactions with 41 species). Although the reduction in reactions was significant, the reduction in species was modest in comparison perhaps due to the fact that RPA is essentially a reaction-lumping technique. Since the computational time scales with the square of the number of species \cite{2008_cnf_pepiot}, alternative methods focusing on reducing the number of species may be more beneficial in this regard. In another study \cite{2015_pcaPlasma_psst_peerenboom}, PCA was used for reducing plasma chemistry but no automated tool was released. Later on, ILDM was reported being used in \cite{2016_jPhysConf_rehman} to accelerate plasma numerical simulations but no standalone chemistry-reduction library was released either. The DRG approach \cite{2005_procCombustInst_lu} was only relatively recently used for reducing plasma chemistry in \cite{2020_psst_sun}. In a more recent work \cite{MAERIVOET2024152006}, sensitivity analysis was used to reduce a large-scale chemical mechanism which was also implemented in a coupled multi-physics simulation to model an atmospheric pressure glow discharge. Despite the aforementioned works, to date, there is a lack of automated reduction libraries for plasma chemistry, and at the same time able to generate chemical mechanisms in the file formats commonly used by the plasma-physics community.

In this work, we present an open-source reduction library tailored to plasma chemistry. In contrast to any currently available tools, \REDLIB\ employs a powerful and well-established reduction method which has long been used for reducing combustion, as well as atmospheric chemical mechanisms \cite{nikolaou_gmd_2018} namely DRGEP: a species-oriented and reaction-oriented reduction approach. The reduction method is controlled by a single parameter, the acceptance threshold, and the user is free to set this for any target species which must be retained in the skeletal chemistry. In section \ref{sec:method}, a brief overview of the DRGEP method is given, section \ref{sec:implementation} gives a detailed description of the software library, and section \ref{sec:validation} demonstrates the use of the library for reducing a large-scale chemical mechanism at various orders. 

\section{The DRGEP method}\label{sec:method}

Consider chemical reactions, $R_i$, 

\begin{equation} \label{eq:chemicalReaction}
R_i:= \sum_{j=1}^{N_s} \nu^{'}_{ij}M_j => \sum_{j=1}^{N_s} \nu^{''}_{ij}M_j
\end{equation}

\noindent where $N_s$ is the total number of species, $M_j$ is the species name, and $\nu^{'}_{ij}$  and $\nu^{''}_{ij}$ are the species stoichiometric coefficients in the reactants and products respectively. In the DRGEP method, a subset chemical mechanism from a detailed set is constructed by removing species which have a negligible effect on their interactions with a predefined set of target species. It is important to note that this is in contrast to other methods such as reaction-lumping \cite{2004_jAtmosphChem_lehman} where essentially a reduced-order model rather than a subset chemical mechanism is created. 

The amount of interaction in DRGEP is quantified by specifying a Direct Interaction Coefficient (DIC), $r_{TB}$, between a target species of interest $T$, and another arbitrary species $B$. Here, we employ the modified DIC as defined in \cite{2008_cnf_pepiot}, 

\begin{equation}\label{eq:dic_coeff}
r_{TB}=\frac{| \sum_{i=1}^{N_r} \dot{w}{_{iT}}\delta_{iB} |}{max(P_T,C_T)}
\end{equation}

\noindent where $N_r$ is the total number of reactions, and $\dot{w}_{iT}$ is the net rate of species $T$ from reaction $i$. The index $\delta_{iB}$ is equal to 1 if $B$ exists in reaction $i$ and 0 otherwise. The net rate of a species $T$ from reaction $i$ is calculated using,

\begin{equation}\label{eq:speciesRate}
\dot{w}_{iT}=\left( \nu^{''}_{iT}-\nu^{'}_{iT} \right) \dot{w}_i
\end{equation}

\noindent where $\dot{w}_i$ is the rate of the reaction, 

\begin{equation}\label{eq:reactionRate}
\dot{w}_i=k_i\prod_{j=1}^{N_{s}}[M_j]^{\nu^{'}_{ij}}
\end{equation}

\noindent where $[M_j]$ is the concentration (or number density for the rate in different units) of species $j$, and $k_i$ is the forward reaction rate constant-ZDPlasKin only considers forward reactions as indicated in Eq. \ref{eq:chemicalReaction}, and reverse reactions (and reverse rate constants) must be given explicitly in the chemical mechanism. The rate constants, $k_i$, may be given in the usual Arrhenius form, in any other user-provided expression, and/or may be calculated on-the-fly by solving for the electron energy distribution function (typically using the BOLSIG+ solver \cite{bolsigSolver}). The production and consumption terms in Eq. \ref{eq:dic_coeff} are given by,

\begin{equation*}
P_T=\sum_{i=1}^{N_r}max(0,\dot{w}_{iT})
\end{equation*}

\begin{equation*}
C_T=\sum_{i=1}^{N_r}max(0,-\dot{w}_{iT})
\end{equation*}

The reaction rates, $\dot{w}_i$, are calculated for a canonical model problem at conditions of interest, and must be provided in a specific binary format for input to \REDLIB . In the utils/ directory of \REDLIB\, we provide a dedicated Fortran module, pcmo.f90, for this purpose.

Clearly, $0 \leq r_{TB}\leq 1$. Large values of $r_{TB}$ indicate that species $B$ is important in determining the rate of $T$, and low values indicate it is less important. The DIC, as defined above, is calculated for all target species defined by the user in a control file. The Path Interaction Coefficient (PIC), $r^p_{TB}$, for a given path $p$ connecting target species $T$ and $B$ is defined as, 

\begin{equation*}
r^p_{TB}=\prod_{i=1}^{N_p-1}r_{M_i}r_{M_{i+1}}
\end{equation*}

\noindent i.e. it is the product of all DICs involving the $N_p$ species along the path $p$ connecting $T$ to $B$. Note that in the above definition $M_1=T$, and $M_{N_p}=B$. The PIC is calculated for all possible paths connecting $T$ to $B$, and an Overall path Interaction Coefficient (OIC), $r^o_{TB}$, is obtained using, 

\begin{equation*}
r^o_{TB}=max(r^p_{TB})
\end{equation*}

The identification of the strongest path is a common problem in computational science, and a number of different route-finding algorithms have been developed throughout the years for this task-the effect of each one on the performance of the DRGEP method has been examined in \cite{2016_cnf_chen}. In this study, we employ a classic, efficient and robust route-finding algorithm for searching through the connected nodes and obtaining $r^o_{TB}$ namely Dijkstra's algorithm  \cite{1959_numerMath_Dijkstra}.

Reduction proceeds by specifying a threshold acceptance interaction value, $a$, and provided $r^o_{TB} \geq a$ species $B$ is considered important and kept in the dependency set, $D_T$, for the target species $T$. In other words, the dependency set for the target is given by,  

\begin{equation*}
D_T=\lbrace M_i: r^o_{TM_i} \geq a \hspace{0.1cm} \forall i=1,2,...,N_s \rbrace
\end{equation*}


\noindent where $N_s$ is the total number of species in the detailed chemical mechanism. Note that for a species $B$ not interacting with $T$, $r^o_{TB}=0$. For a number, $N_{target}$, of target species  defined by the user using their corresponding indices, ${j_1, j_2,..., j_{N_{target}}}$ in the species list of the chemical mechanism, the final skeletal species set, $S^{skel}$ , is obtained by the union of all target species dependency sets, 

\begin{equation*}
S^{skel}=D_{M_{j_1}} \cup D_{M_{j_2}} \cup ... \cup D_{M_{j_{N_{target}}}}
\end{equation*}

For larger values of the acceptance threshold $a$, more species are rejected while for lower values more species are accepted in each target's dependency set, and the skeletal mechanism grows larger. This parameter allows the user to control the amount of reduction. \REDLIB\ also allows the user to set different values of the acceptance threshold for each target species. The number of species in the set $S^{skel}$ is given by its cardinality i.e. $N_s^{skel}=|S^{skel}|$. Finally, only reactions involving species in $S^{skel}$ are retained, and the remaining reactions removed. Specifically, the reduced reaction set, $R^{skel}$, is given by, 

\begin{equation*}
R^{skel}=\lbrace R_i \hspace{0.1cm} \forall i: M_{ij} \hspace{0.1cm} \in S^{skel} \hspace{0.1cm} \forall j \in 1,2, ..., N^i_s \rbrace
\end{equation*}

\noindent where $N^i_s$ is the number of (distinct) species taking part in reaction $i$. In practice, in the current implementation of the method, OICs are calculated for all data points of the canonical problem.

\section{Implementation}\label{sec:implementation}

\begin{figure}[h!]
\centering
\includegraphics[scale=0.75, trim={4cm 4cm 2cm 2cm}]{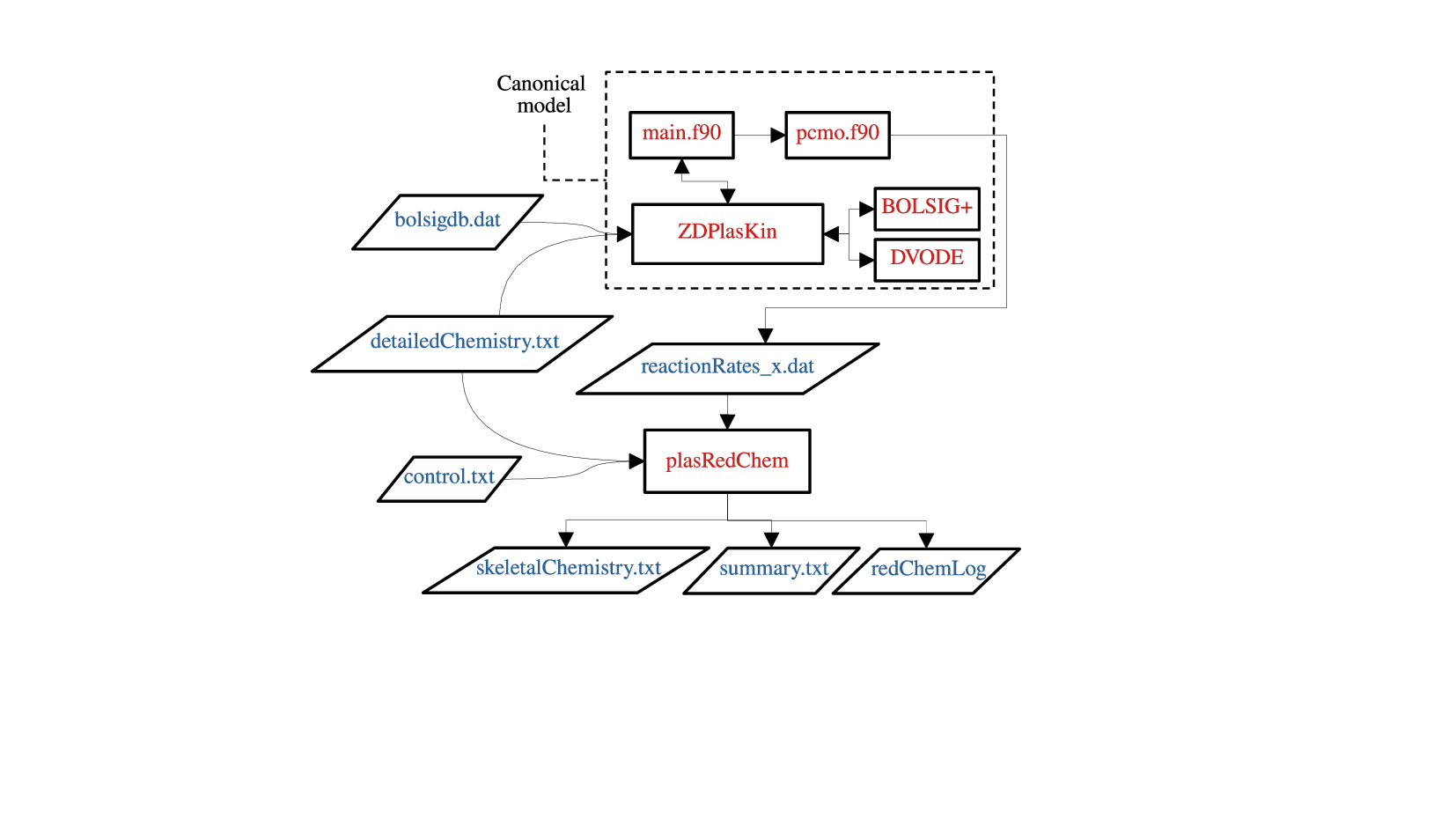}
\caption{Workflow of \REDLIB: inputs consist of a detailed chemical mechanism in ZDPlasKin format \cite{zdplaskin}, reaction-rate data (binary format), and an input reduction-control file. Module pcmo.f90 can be included by the user in the main.f90 file to write the data in the format required by \REDLIB.} 
\label{fig:workflow}
\end{figure}

Figure \ref{fig:workflow} shows the workflow of \REDLIB. 
The inputs to \REDLIB\ consist of a chemical mechanism given in ZDPlasKin format containing all considered species and reactions (including electron impact processes), as well as the corresponding rate coefficients. \REDLIB\ also requires a control file where the user sets the reduction parameters, and reaction rate data obtained by solving a canonical model problem. Any canonical model may be used (0D, 1D etc.). In this work, we have used ZDPlasKin \cite{zdplaskin} which solves the governing equations for a 0D initial-value problem-the model is described in detail in section \ref{sec:canonicalModel}. ZDPlasKin requires a separate file, bolsigdb.dat, which contains the cross-section data for the electron collision reactions. ZDPlasKin (written in Fortran 90) tracks the time evolution of user-defined species and a spatially uniform gas temperature. One of its components is a stand-alone pre-processor which first converts the text-based chemical mechanism (detailedChemistry.txt) into a bespoke Fortran module. The user provides a main.f90 file in which he/she sets the initial conditions and calls ZDPlasKin native routines. At every time step, ZDPlasKin: (i) requests electron-impact rate coefficients and transport data from the embedded Boltzmann solver BOLSIG+ \cite{bolsigSolver} by solving for the instantaneous Electron-Energy-Distribution Function (EEDF) at the current reduced electric field, (ii) integrates the resulting stiff system of differential equations with the DVODE solver, and (iii) returns the updated number densities, gas temperature, and other user-defined outputs. 

The reaction rate data needed by \REDLIB\ are in binary Fortran format-we provide a module, pcmo.f90, which the user can easily import into main.f90 and write these to file. In the control file, the user sets the name of the chemical mechanism, the number of different reaction rate cases and the number of data-points to use, the number of target species $T$ and their indices in the detailed chemical mechanism, and finally the acceptance threshold for each target species. In the ``examples" directory of the library we provide a full working example along with reaction rate data for the user. 

Once the user runs the reduction library, a skeletal chemical mechanism in ZDPlasKin format is generated automatically, and written in the ``output" directory. A log of the numerical computations is also written as well as a summary of the reduction statistics and parameters. In the summary file, a ranking of the DICs for each target species is given which the user may use to identify the dominant species in the dependency set. 

\section{Validation}\label{sec:validation}

\subsection{Canonical model}\label{sec:canonicalModel}

\begin{figure}[h!]
\subfigure[]{
\includegraphics[width=0.50\textwidth, trim=2.0cm 0.0cm 0.0cm 0.0cm]{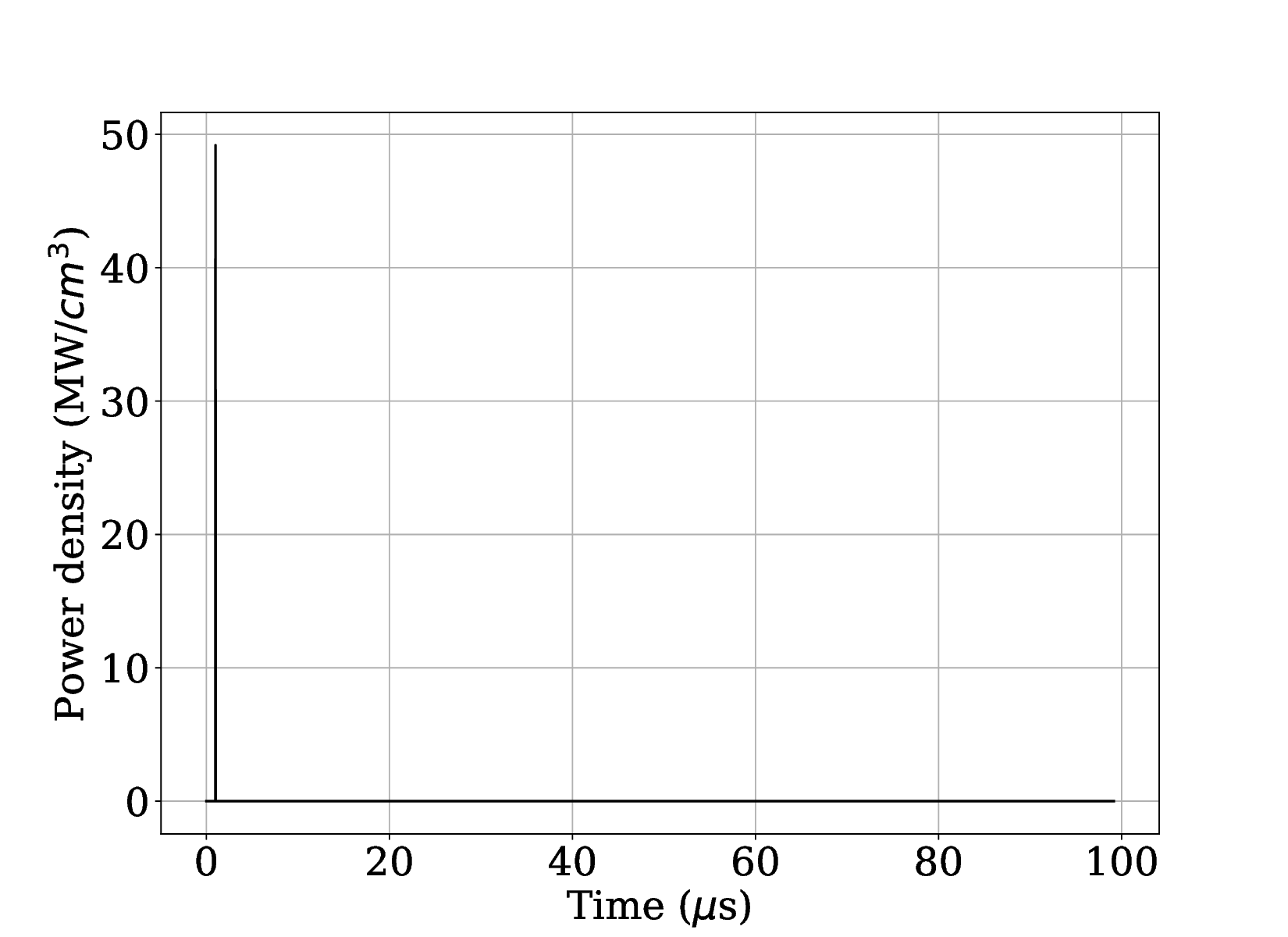}
}
\subfigure[]{
\includegraphics[width=0.50\textwidth, trim=2.0cm 0.0cm 0.0cm 0.0cm]{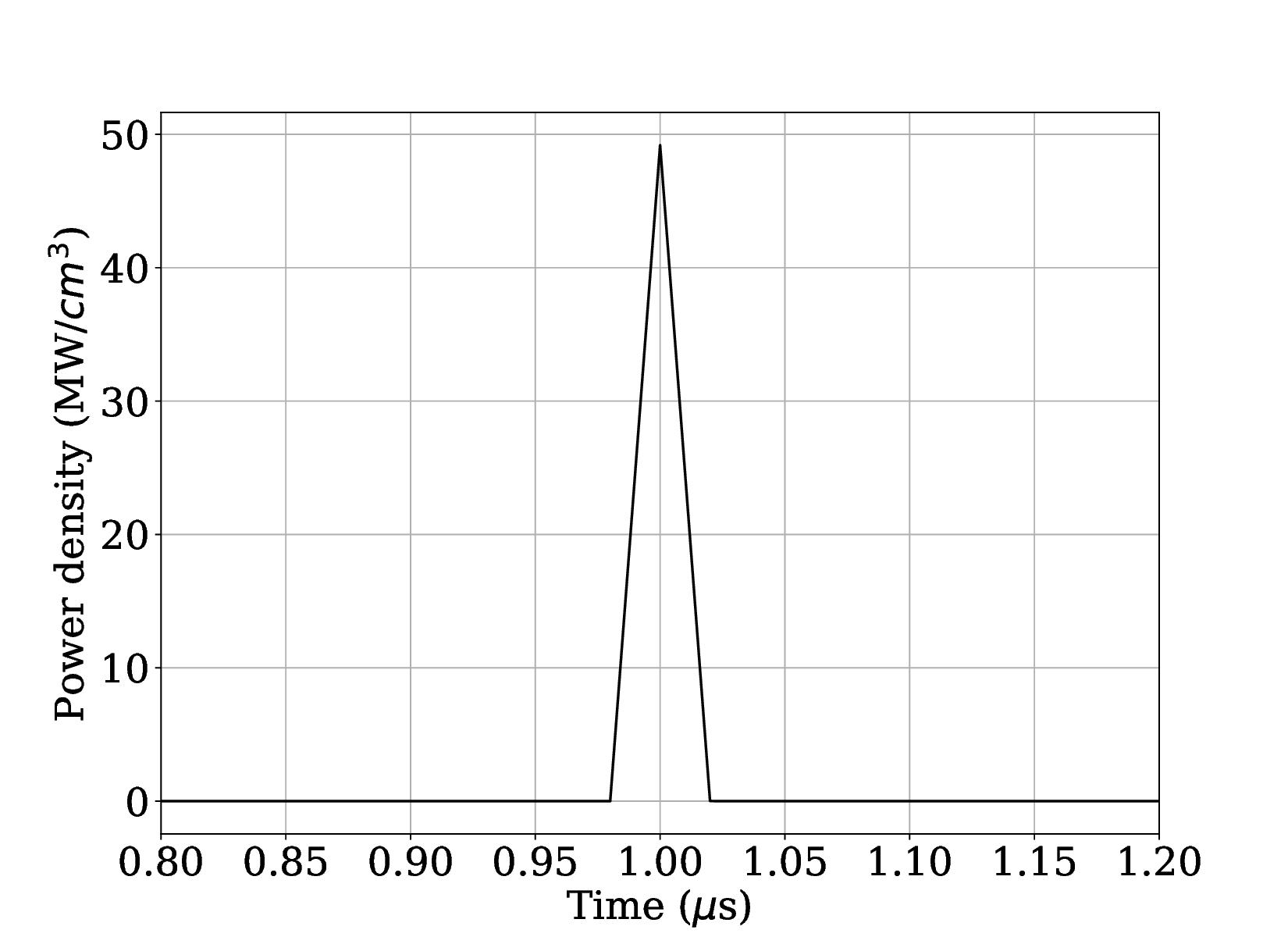}
}
\caption{(a) The applied power pulse profile, and (b) its close-up. Peak power density is 49.2 MW/$\mathrm{cm^3}$, and the pulse width is 40 ns.}\label{fig:powerProfile}
\end{figure}

The canonical model used to generate the reaction rate data is similar to the 0D model developed in \cite{2023_chemEngJourn_morais,2024_ppp_morais}, and which has been validated against experimental data. A Nano-second Pulsed Discharge (NPD) plasma is used to reform methane into hydrogen and other useful components. A constant-volume reactor filled initially with an inert mixture of methane and traces of hydrogen is subjected to a power pulse profile as shown in Fig. \ref{fig:powerProfile}. Upon power striking, a resulting peak in the electric field occurs. In turn, this accelerates free electrons which collide with the methane and hydrogen molecules, leading to various ionisation (increasing the density of free  electrons), dissociation and excitation processes. This creates the ions, radicals, excited species and stable molecules characteristic of methane plasmas. The profile of the pulse is selected to represent laboratory conditions by specifying the peak power, the width of the pulse, and the plasma volume within the reactor \cite{2023_chemEngJourn_morais}.

ZDPlasKin integrates the species-balance equations for the number densities of every neutral, radical, ion or electronically/vibrationally excited state defined by the user in the chemical mechanism,

\begin{equation}\label{eq:rateNoDensity}
\ddt{n_k}=\sum_{r=1}^{N_r} \dot{w}_{kr}
\end{equation}

\noindent subject to $n_k(t=0)=n^0_k$ where $n_k$ is the number density of species $k$, and $\dot{w}_{kr}$ is the species rate (no of species per unit volume per unit time) of species $k$ from reaction $r$. The sum in Eq. \ref{eq:rateNoDensity} is all over the $N_r$ reactions in the chemical mechanism. The species rate from every reaction, $\dot{w}_{kr}$, is obtained using Eqs. \ref{eq:speciesRate} and \ref{eq:reactionRate}. The initial number density of the electrons $n^0_e=1.0 \cdot 10 ^{11}$, the initial gas temperature $T^0_g=293.15$ K, and the pressure $p=111457.5$ Pa.  For heavy-particle reactions, tabulated or Arrhenius-type coefficients taken from literature are read directly by the solver to calculate ${k_r}$ in Eq. \ref{eq:reactionRate}. For electron-impact reactions, the rate constants are calculated on-the-fly: ZDPlasKin calls the Boltzmann solver BOLSIG+ to calculate the electron energy distribution function from an imposed $E/N$, and the corresponding cross-sections are then integrated to yield the rate constant.

The electric field, $E$, changes in time as a result of the applied power profile, $p(t)$, and is calculated using the Joule relation,

\begin{equation*}
\frac{E(t)}{n_g}=\frac{1}{n_g}\sqrt{\frac{p(t)}{\sigma}}
\end{equation*}

\noindent where the electron conductivity $\sigma=n_e\mu_ee$, $\mu_e$ is the electron mobility coefficient, and $n_g$ is the gas number density. 

A self-consistent thermal evaluation is available in ZDPlasKin, in which a spatially uniform gas temperature is calculated in time by solving the constant-volume energy balance, 

\begin{equation*}
\frac{n_g}{\gamma-1}\ddt{T_g}=\sum_{r=1}^{Nr} \dot{w}_r \delta \epsilon _r + n_e P_{elas}-\dot{Q}
\end{equation*}

\noindent where $\delta \epsilon _r$ is the energy released/consumed in reaction $r$, and $P_{elas}$ is the energy gained by the gas due to elastic collisions with the electrons. The source term $\dot{Q}$ represents heat exchanges with the environment. In line with \cite{2024_ppp_morais}, energy exchanges between the reactor and the environment are modelled using the expression, 

\begin{equation*}
\dot{Q}=8\frac{\lambda_{CH4}}{r^2}\left( T_{g}-T_{e} \right)
\end{equation*}

\noindent where $\lambda_{CH_4}$ is the thermal conductivity of methane, $r$ is the radius of the plasma zone, $T_{g}$ is the gas temperature, and $T_e$ is the environment temperature (standard conditions). The thermal conductivity is modelled using a linear relationship \cite{2024_ppp_morais}, 

\begin{equation*}
\lambda_{CH_4}=1.48\cdot 10^{-6} T_g-9.92\cdot 10^{-5}
\end{equation*}

For further details on the model and its implementation we refer the reader to the works in \cite{2023_chemEngJourn_morais,2024_ppp_morais}.

\subsection{Chemical mechanism}

In order to validate \REDLIB, a detailed chemical mechanism describing methane-hydrogen plasmas is considered. It contains 77 species and 4404 reactions comprising ground state stable molecules, vibrationally and electronically excited states, as well as the corresponding radicals and ions. The current mechanism is built upon previous chemistry sets developed specifically for plasma-based methane conversion [52, 53]. These were extensively validated against experimental data from NPD plasmas in a range of pressures, gas temperatures and power input, which demonstrates the ability of this chemical mechanism to correctly capture the gas-phase kinetics in $\mathrm{CH_4/H_2}$ plasmas. Table \ref{tbl:chemMechSpecies} gives a detailed breakdown of the different species in the chemical mechanism.

\begin{table}[h!]
\centering
\caption{The 77 species in the methane-hydrogen detailed plasma chemical mechanism: 18 charged species, 19 excited species, and 40 stable molecules and radicals.}
\footnotesize
\begin{tabular}{lll}
\hline
Ions and electrons & Excited molecules  & Radicals and stable molecules \\
\hline
\\                                                                                                                                                          
e, $\mathrm{H^-}$,  $\mathrm{C^+}$, $\mathrm{C_2^+}$, $\mathrm{C_2H_6^+}$, & Electronic: H(E2)         & $\mathrm{H_2}$, H \\
$\mathrm{C_2H_5^+}$, $\mathrm{C_2H_4^+}$, $\mathrm{C_2H_3^+}$, & Vibrational: $\mathrm{CH_4(V1)}$,  & $\mathrm{C_1}$: $\mathrm{CH_4}$, $\mathrm{CH_3}$, $\mathrm{CH_2}$, \\
$\mathrm{C_2H_2^+}$, $\mathrm{C_2H^+}$, $\mathrm{CH_5^+}$,   &  $\mathrm{CH_4(V2)}$, $\mathrm{CH_4(V3)}$,       & CH, C(S) \\
$\mathrm{CH_4^+}$, $\mathrm{CH_3^+}$, $\mathrm{CH_2^+}$,    & $\mathrm{CH_4(V4)}$, $\mathrm{H_2(V1)}$,          & $\mathrm{C_2}$: $\mathrm{C_2H_6}$, $\mathrm{C_2H_4}$, $\mathrm{C_2H_2}$,\\
$\mathrm{CH^+}$, $\mathrm{H_3^+}$, $\mathrm{H_2^+}$, $\mathrm{H^+}$ & $\mathrm{H_2(V2)}$, $\mathrm{H_2(V3)}$,          & $\mathrm{C_2H_5}$, $\mathrm{C_2H_3, C_2H}$, \\ 
                              & $\mathrm{H_2(V4)}$, $\mathrm{H_2(V5)}$,                   & $\mathrm{C_3}$: $\mathrm{CH_3CH_2CH_3}$, \\
                              & $\mathrm{H_2(V6)}$, $\mathrm{H_2(V7)}$,                                 & $\mathrm{CH_3CH_2CH_2}$, $\mathrm{CH_3CHCH_3}$, \\
                              & $\mathrm{H_2(V8)}$, $\mathrm{H_2(V9)}$                                & $\mathrm{CH_2CHCH_3}$, $\mathrm{CH_2CHCH_2}$, \\
                              & $\mathrm{H_2(V10)}$, $\mathrm{H_2(V11)}$,                                 & $\mathrm{CHCHCH_3}$, $\mathrm{CH2CCH_3}$, \\
                              & $\mathrm{H_2(V12)}$, $\mathrm{H_2(V13)}$,                                & $\mathrm{CH_3CCH}$, $\mathrm{H2CCCH_2}$, \\
                              &  $\mathrm{H_2(V14)}$                                & $\mathrm{HCCCH_2}$, $\mathrm{HCCCH}$, \\
                              &                                 & $\mathrm{C_4}$:  $\mathrm{CH_3CH_2CH_2CH_3}$, \\
                              &                                 & $\mathrm{CH_2CH_2CH_2CH_3}$, \\
                              &                                 & $\mathrm{CH_3CHCH_2CH_3}$, \\
                              &                                 & $\mathrm{CH_2CHCH_2CH_3}$, \\
                              &                                 & $\mathrm{CH_3CHCHCH_3}$, \\
                              &                                 & $\mathrm{C_4H_7}$, $\mathrm{CH_2CHCH_2CH_2}$, \\
                              &                                 & $\mathrm{CH_2CHCHCH_3}$, \\
                              &                                 & $\mathrm{C_4H_6, C_4H_5}$,  \\
                              &                                 & $\mathrm{HCCCHCH_2}$, $\mathrm{C_4H_3}$, $\mathrm{HCCCCH}$ \\
                              &                                 & Carbon species: $\mathrm{C_3}$, $\mathrm{C_2}$, $\mathrm{C}$ \\ 
\hline
\end{tabular}
\label{tbl:chemMechSpecies}
\end{table}

\subsection{Reduction results}

After solving the governing equations as described above using the detailed chemical mechanism, the corresponding reaction rate data are saved, and \REDLIB\ is used to generate a number of different skeletal mechanisms. Each one is generated by varying the acceptance threshold. For targets, we select 9 species as shown in Table \ref{tbl:reductionParams}. The choice of target species is up to the user. In this study, we chose a set of species including electrons, the hydrogen and methane molecules as well as a few vibrationally-excited states for these molecules in order to illustrate the use of the library. 

\begin{table}[ht!]
\centering
\caption{Reduction parameters: the 9 target species, the number of reaction rate datasets used for the NPD plasma, and the range of acceptance thresholds.}
\footnotesize
\begin{tabular}{lll}
\hline
Target species & No data sets  & Acceptance threshold \\
\hline
\\                                                                                                                                                          
e, $\mathrm{H_2}$, $\mathrm{CH_4}$, $\mathrm{CH_4(V1)}$, $\mathrm{CH_4(V2)}$,  & 738 & $0-0.8$ \\
$\mathrm{CH_4(V3)}$, $\mathrm{CH_4(V4)}$, $\mathrm{H_2(V1)}$, $\mathrm{H_2(V2)}$ & & \\
\hline
\end{tabular}
\label{tbl:reductionParams}
\end{table}

\begin{figure}[h!]
\centering
\includegraphics[scale=0.3, trim=2.0cm 0.0cm 0.0cm 0.0cm]{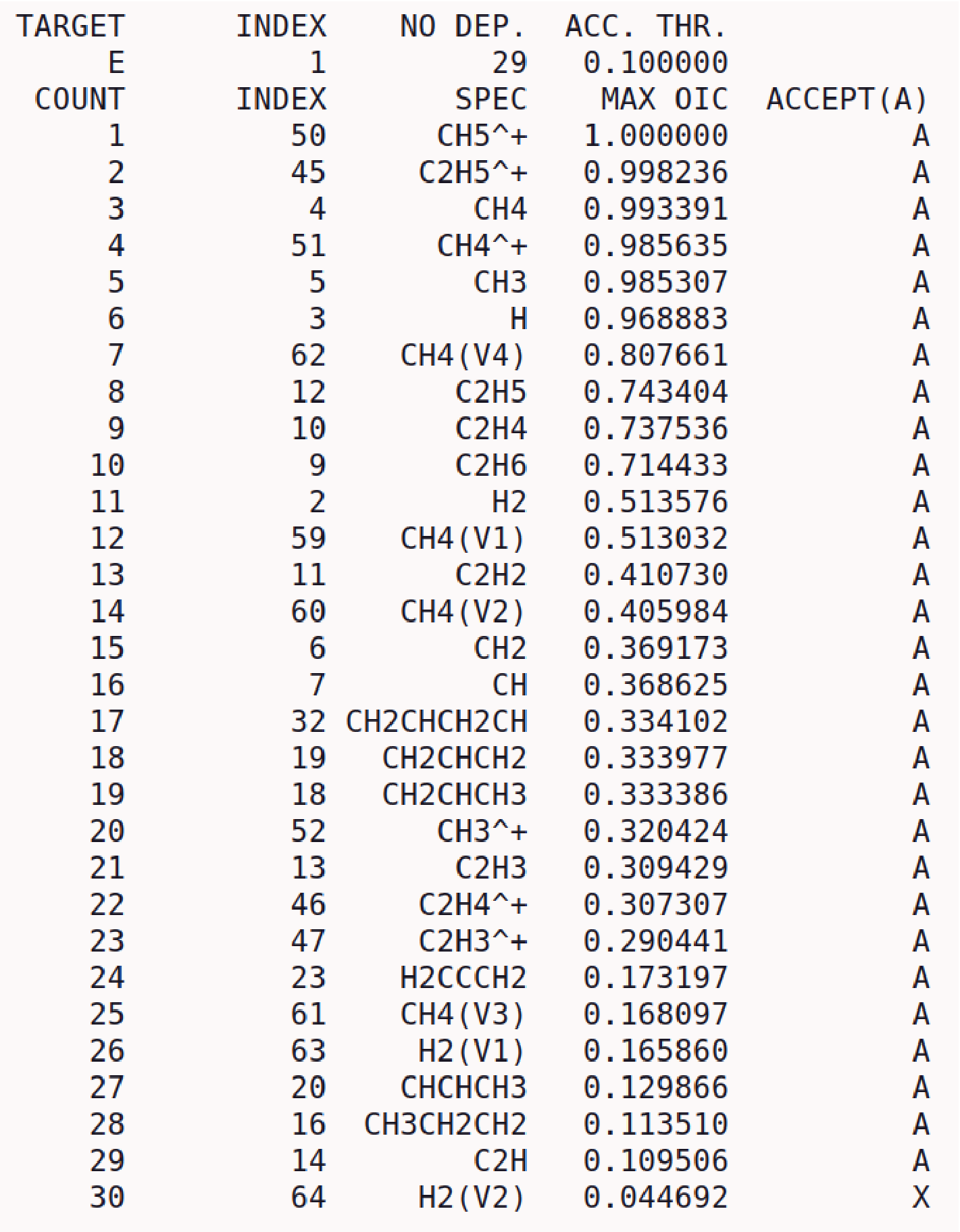}
\caption{A snapshot of the summary file (summary.txt) after running \REDLIB\ (in this case for skeletal mechanism A) showing the ranking of species for target E (electrons) based on their OIC: the acceptance threshold is 0.1, and 29 species are retained in the skeletal mechanism.}
\label{fig:reductionSummary}
\end{figure}

Figure \ref{fig:reductionSummary} shows an example of the reduction summary file after running \REDLIB. This file contains information on the target species, their indices in the detailed chemical mechanism list, the number of dependencies i.e. the cardinality of the dependency set for each target, and the acceptance threshold. Also shown, is the ranking of the species in the dependency set based on their maximum OIC as obtained by running through the different data sets. For this particular case, we see from Fig. \ref{fig:reductionSummary} that the most dominant species for the electron target species is $\mathrm{CH_5^+}$ followed by $\mathrm{C_2H_5^+}$. For an acceptance threshold of 0.1, 29 species are retained down to $\mathrm{C_2H}$ after which the OIC falls sharply below the acceptance threshold. This summary file is very helpful to the user as it can help to shed light into the dependencies for each target. 

\begin{figure}[h!]
\subfigure[]{
\includegraphics[width=0.50\textwidth, trim=2.0cm 0.0cm 0.0cm 0.0cm]{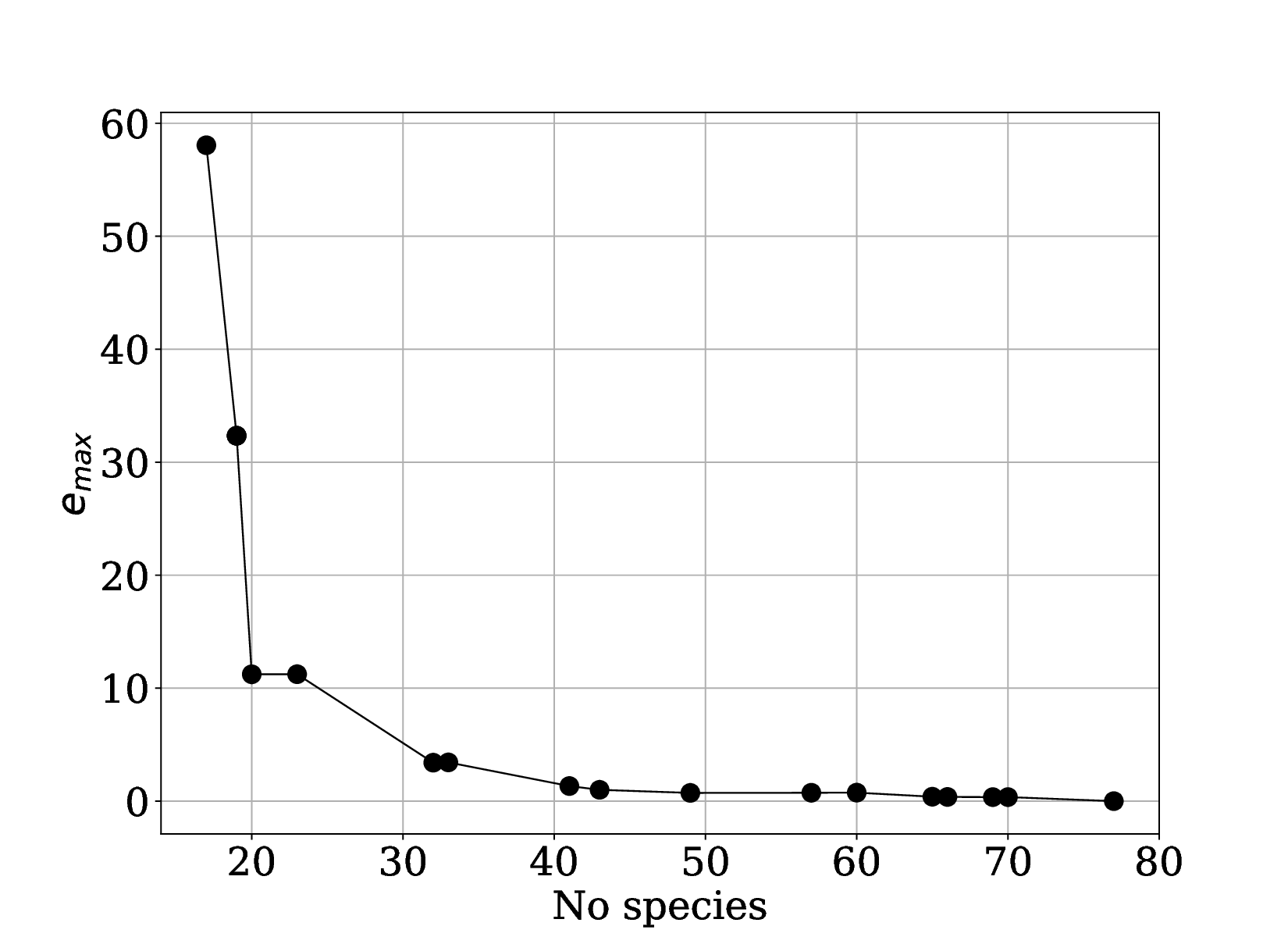}
}
\subfigure[]{
\includegraphics[width=0.50\textwidth, trim=2.0cm 0.0cm 0.0cm 0.0cm]{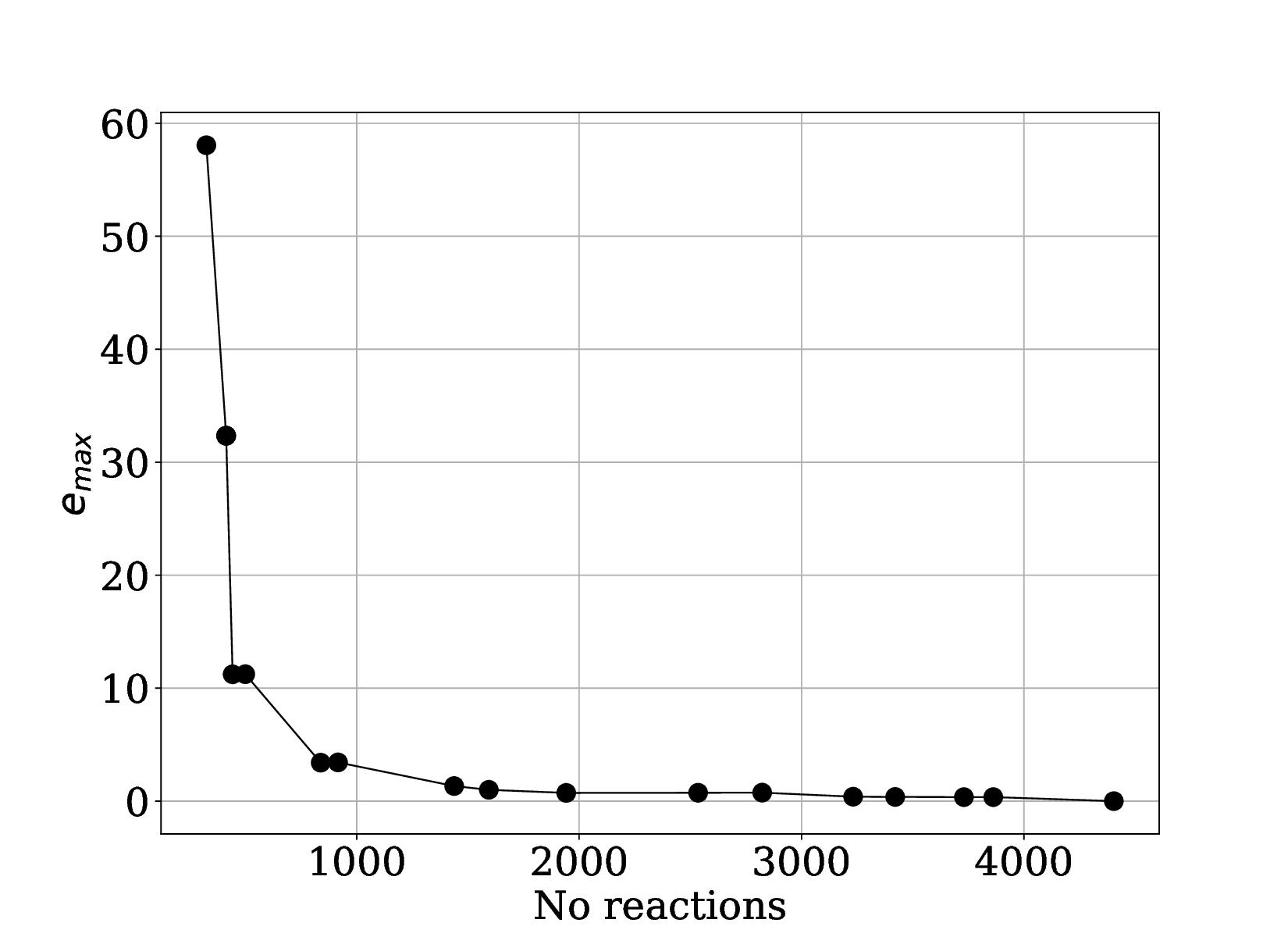}
}
\caption{$e_{max}$ (Eq. \ref{eq:meanMaxError}) against number of species (a), and number of reactions (b) in the generated skeletal mechanisms.}
\label{fig:percError}
\end{figure}

In order to quantify the effect of reduction on the accuracy of the resulting skeletal mechanism, an instantaneous error measure, $e_T(t)$, is defined for each target species as follows, 

\begin{equation}\label{eq:localError}
e_T(t)=\frac{100 \cdot |n^{skel}_{T}(t)-n^{det}_{T}(t)|}{|n^{det}_{T}(t)|}
\end{equation}

\noindent where $n^{skel}_T$ is the number density profile for target species $T$ obtained using the skeletal chemistry, and $n^{det}_T$ using the detailed chemistry. For all targets, we define a mean maximum error,

\begin{equation}\label{eq:meanMaxError}
e_{max}=\frac{1}{N_{target}}\sum_{i=1}^{N_{target}}max(e_i(t))
\end{equation}

\noindent i.e. it is the mean of the maximum absolute percentage difference for all target species. Figure \ref{fig:percError}(a) shows $e_{max}$ against the number of species in each generated skeletal mechanism. Also shown, in Fig. \ref{fig:percError}(b) is the corresponding number of reactions. With an increasing acceptance threshold, the number of species/reactions reduces, and the corresponding error increases. We observe from Fig. \ref{fig:percError} (a) that the error remains well below 5\% down to about 32 species and 838 reactions while still obtaining significant reduction. At 32 species, $e_{max}=3.4\%$ only while the species reduction factor is 2.4 and the reactions reduction factor is 5.3 both of which present a substantial reduction.  

\begin{table}[ht!]
\centering
\caption{The species retained in skeletal mechanisms A and B: 32 for A and 20 for B.}
\footnotesize
\begin{tabular}{cl}
\hline
Skeletal mechanism & Species  \\
\hline
\\                                                                                                                                                          
A & $\mathrm{e}$, $\mathrm{H_2}$, $\mathrm{H}$, $\mathrm{CH_4}$, $\mathrm{CH_3}$, $\mathrm{CH_2}$, $\mathrm{CH}$, $\mathrm{C_2H_6}$, $\mathrm{C_2H_4}$, $\mathrm{C_2H_2}$, \\
  & $\mathrm{C_2H_5}$, $\mathrm{C_2H_3}$, $\mathrm{C_2H}$, $\mathrm{CH_3CH_2CH_2}$, $\mathrm{CH_2CHCH_3}$, \\
  & $\mathrm{CH_2CHCH_2}$, $\mathrm{CHCHCH_3}$, $\mathrm{H_2CCCH_2}$, $\mathrm{CH_2CHCH_2CH_2}$,  \\
  & $\mathrm{C_2H_5^+}$, $\mathrm{C_2H_4^+}$, $\mathrm{C_2H_3^+}$, $\mathrm{CH_5^+}$, $\mathrm{CH_4^+}$, $\mathrm{CH_3^+}$, \\ 
  & $\mathrm{CH_4(V1)}$, $\mathrm{CH_4(V2)}$, $\mathrm{CH_4(V3)}$, $\mathrm{CH_4(V4)}$, \\
  & $\mathrm{H_2(V1)}$, $\mathrm{H_2(V2)}$, $\mathrm{H_2(V3)}$     	  \\
B & $\mathrm{e}$, $\mathrm{H_2}$, $\mathrm{H}$, $\mathrm{CH_4}$, $\mathrm{CH_3}$, $\mathrm{CH_2}$, $\mathrm{CH}$, $\mathrm{C_2H_6}$, $\mathrm{C_2H_4}$, $\mathrm{C_2H_2}$, \\
  & $\mathrm{C_2H_5}$, $\mathrm{C_2H_5^+}$, $\mathrm{CH_5^+}$, $\mathrm{CH_4^+}$, $\mathrm{CH_4(V1)}$, $\mathrm{CH_4(V2)}$, $\mathrm{CH_4(V3)}$, \\ 
  & $\mathrm{CH_4(V4)}$, $\mathrm{H_2(V1)}$, $\mathrm{H_2(V2)}$ \\
\hline
\end{tabular}
\label{tbl:speciesForSkelAB}
\end{table}

In order to test the performance of the skeletal mechanisms generated using \REDLIB, we compare the number density profiles obtained using the skeletal mechanisms with those obtained using the detailed one. This is done at two rather large reduction levels in order to present a stringent test case: one at 32 species and 838 reactions (Skeletal A), and another one at only 20 species and 442 reactions where $e_{max}=11.2\%$ (Skeletal B). Table \ref{tbl:speciesForSkelAB} shows the different species retained for the generated skeletal mechanisms A and B. 

\begin{figure}[h!]
\subfigure[]{
\includegraphics[width=0.50\textwidth, trim=2.0cm 0.0cm 0.0cm 0.0cm]{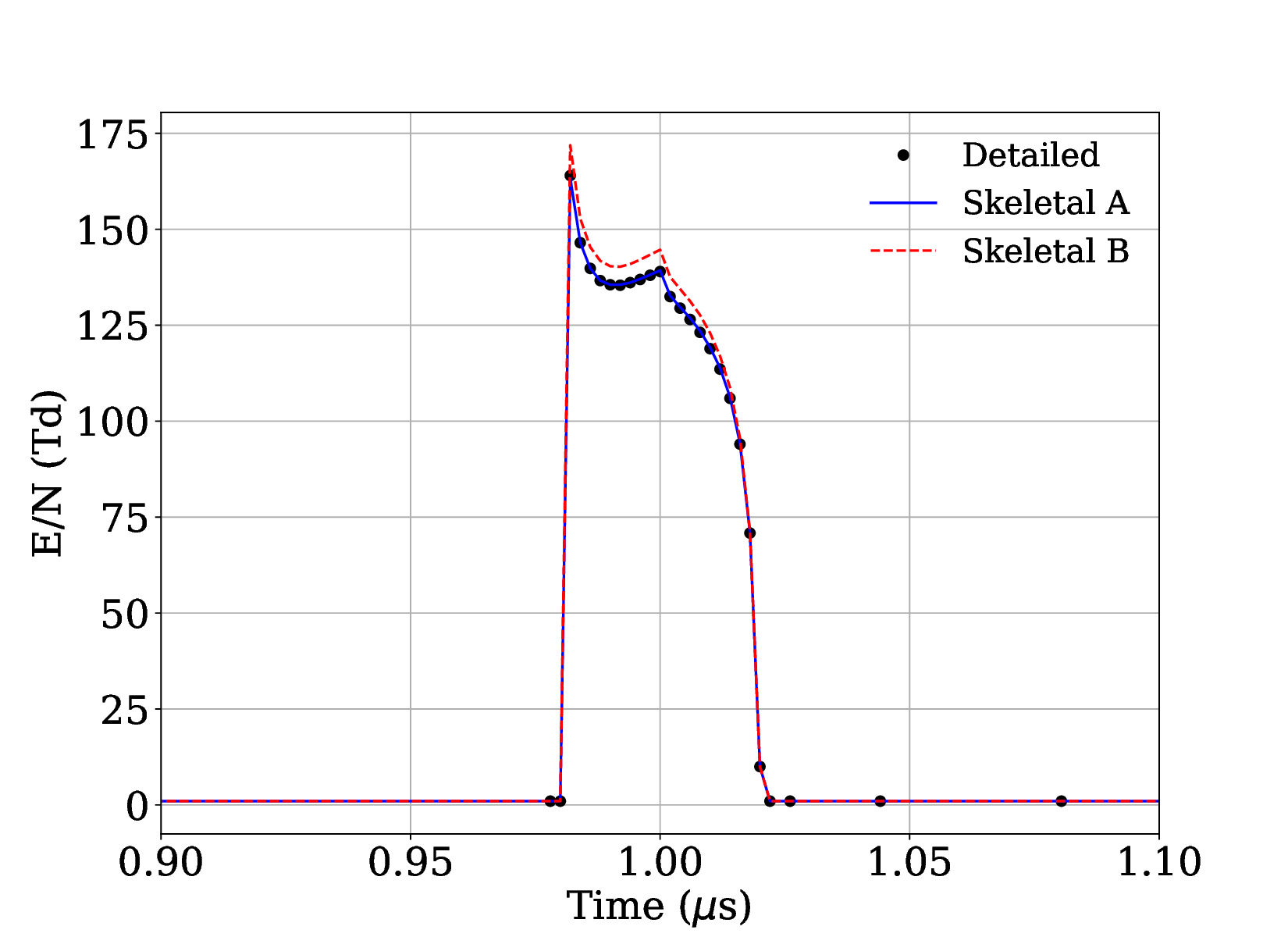}
}
\subfigure[]{
\includegraphics[width=0.50\textwidth, trim=2.0cm 0.0cm 0.0cm 0.0cm]{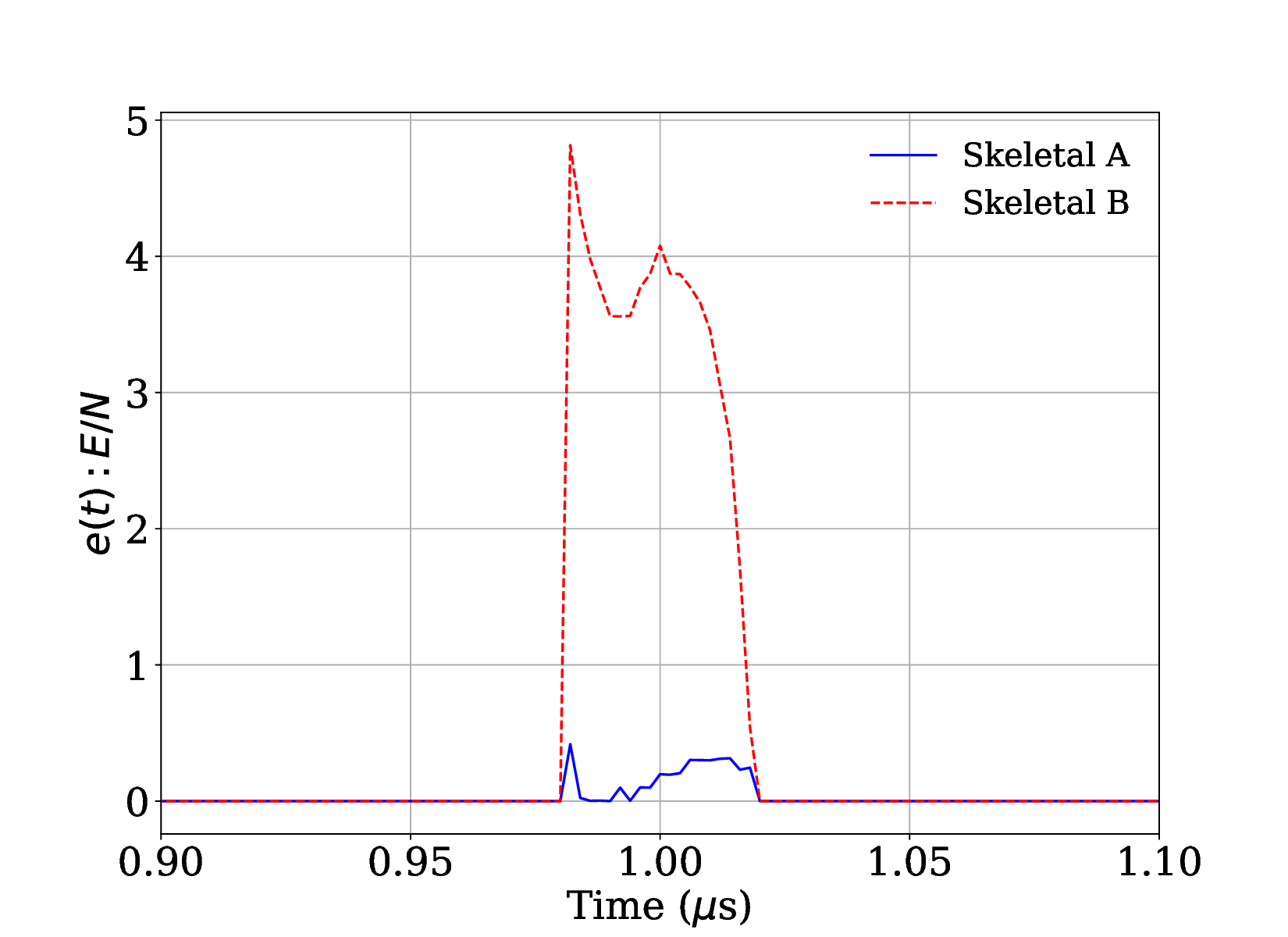}
}
\subfigure[]{
\includegraphics[width=0.50\textwidth, trim=2.0cm 0.0cm 0.0cm 0.0cm]{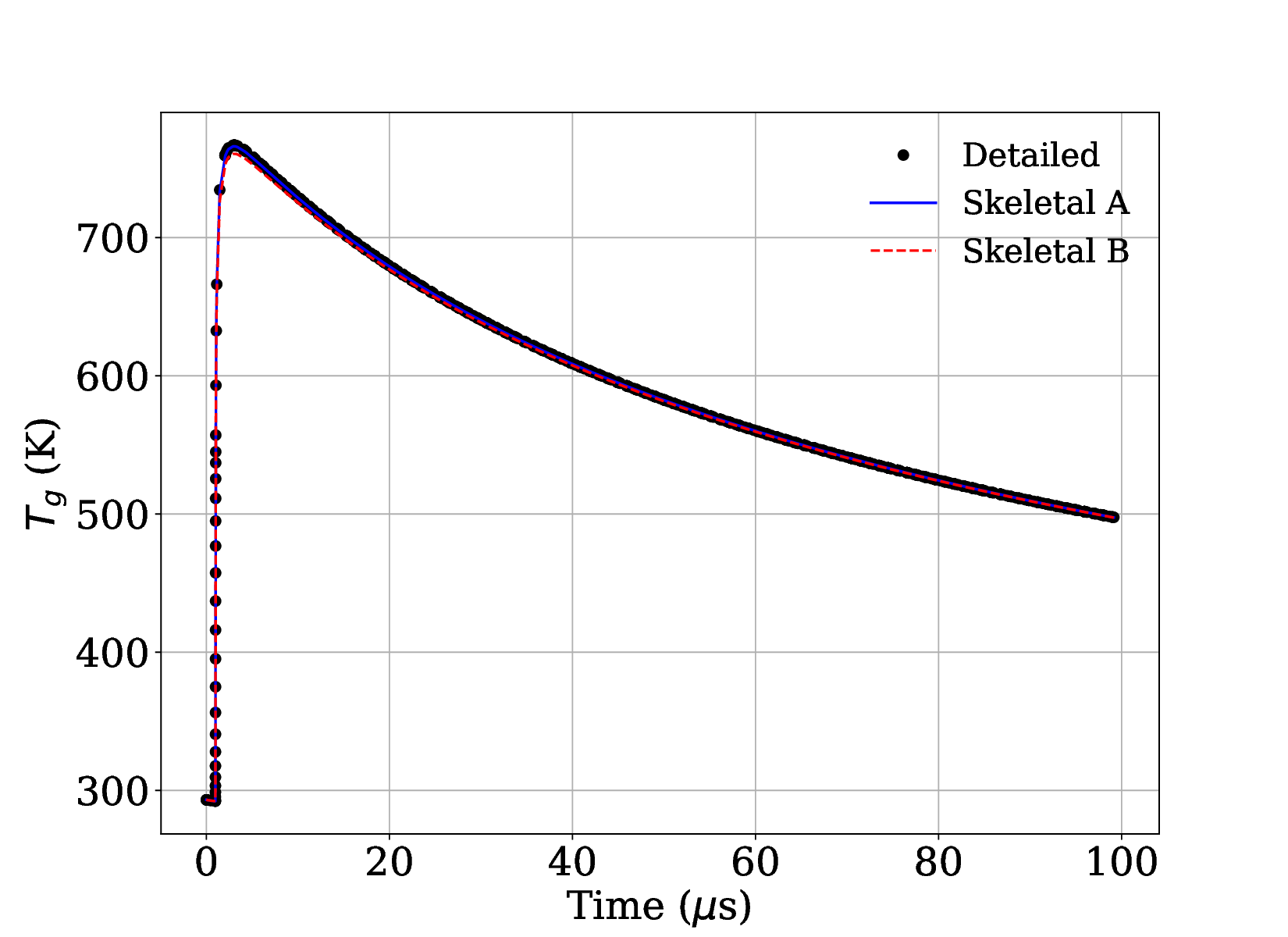}
}
\subfigure[]{
\includegraphics[width=0.50\textwidth, trim=2.0cm 0.0cm 0.0cm 0.0cm]{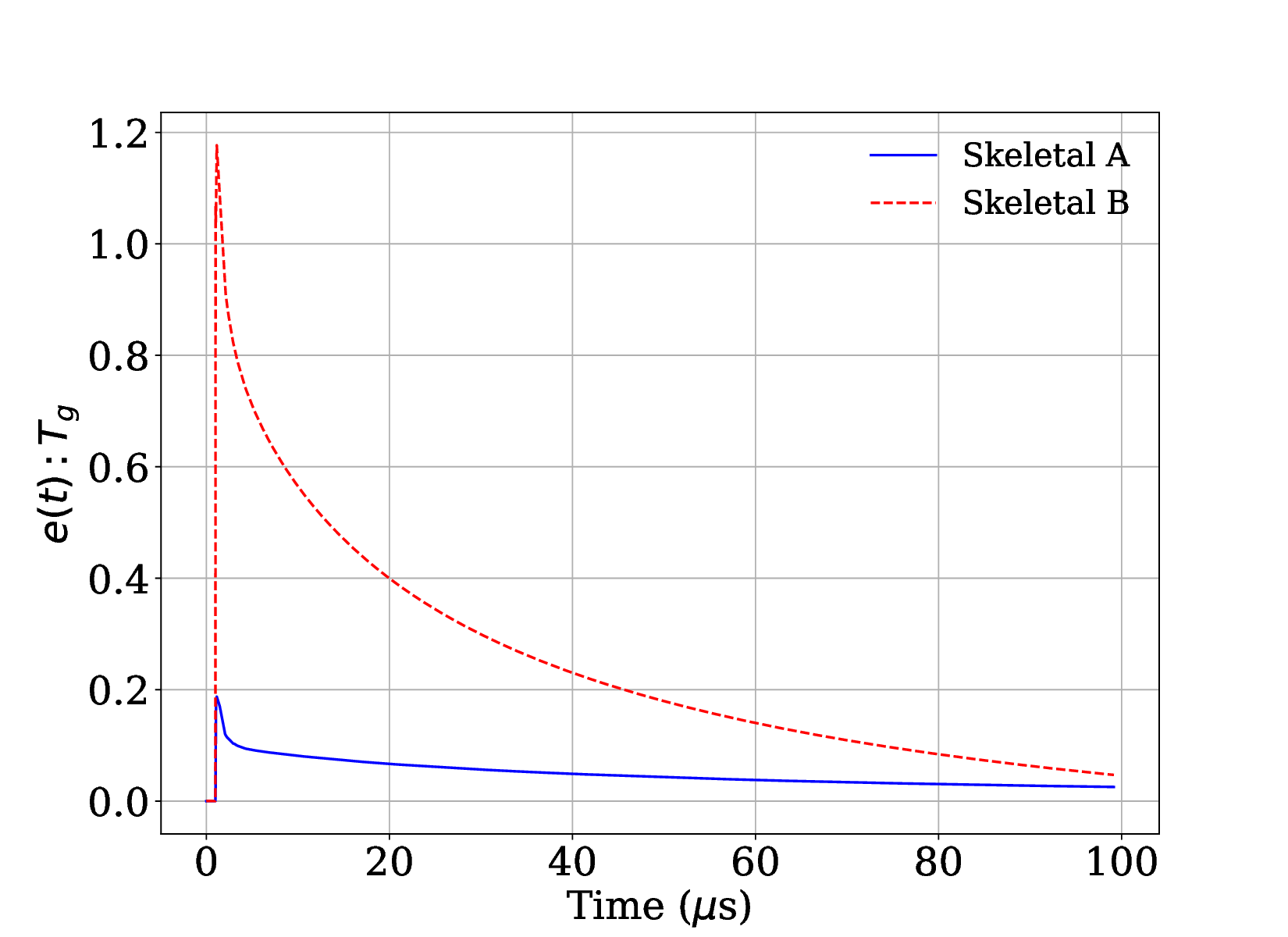}
}
\caption{Reduced electric field, temperature, and associated errors for each, $e(t)$ (Eq. \ref{eq:localError}), obtained using the detailed chemical mechanism (77 species), and two skeletal mechanisms: A (32 species), and B (20 species).}
\label{fig:skelVsDetEandT}
\end{figure}

\begin{figure}[h!]
\subfigure[]{
\includegraphics[width=0.50\textwidth, trim=2.0cm 0.0cm 0.0cm 0.0cm]{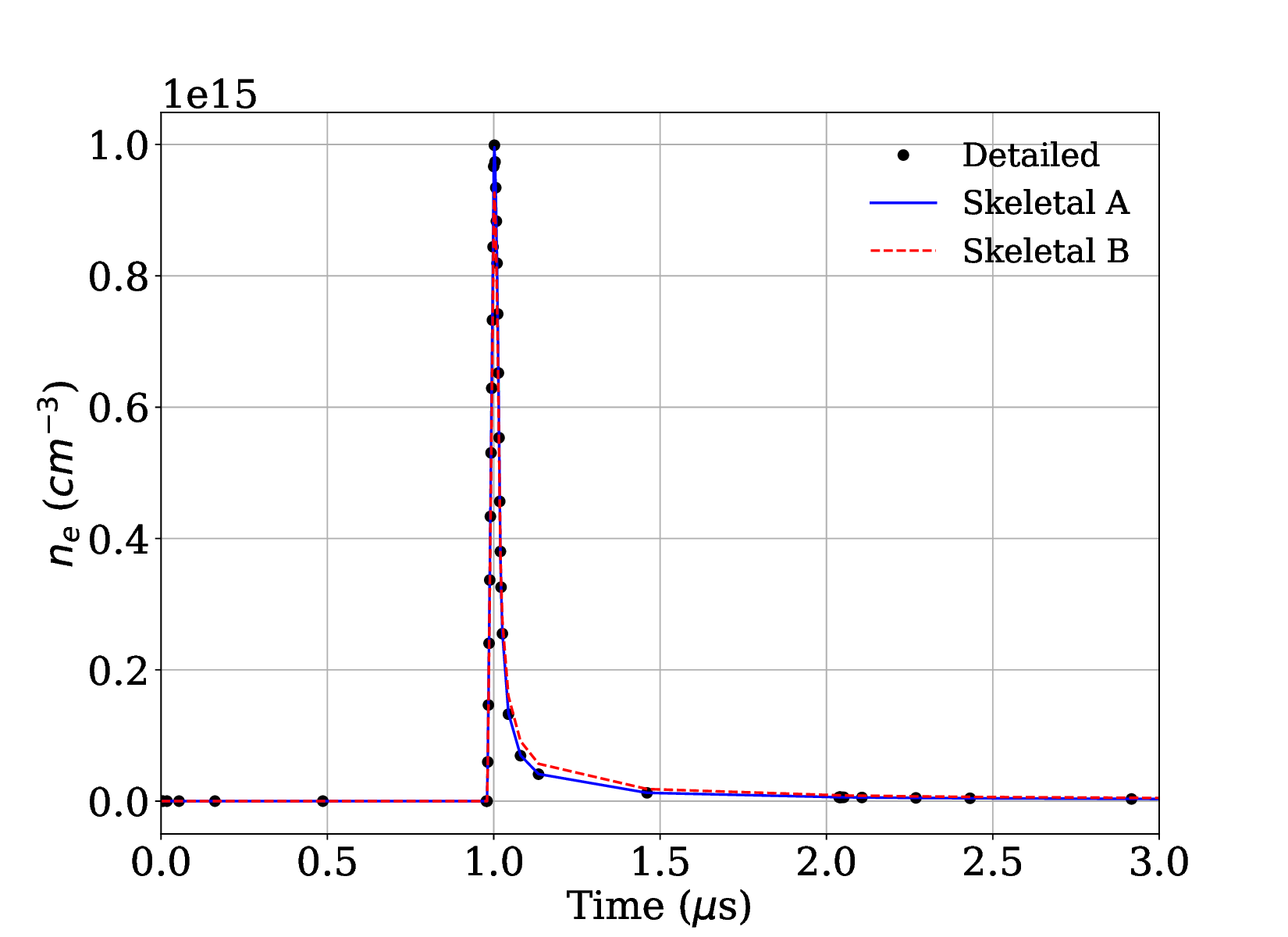}
}
\subfigure[]{
\includegraphics[width=0.50\textwidth, trim=2.0cm 0.0cm 0.0cm 0.0cm]{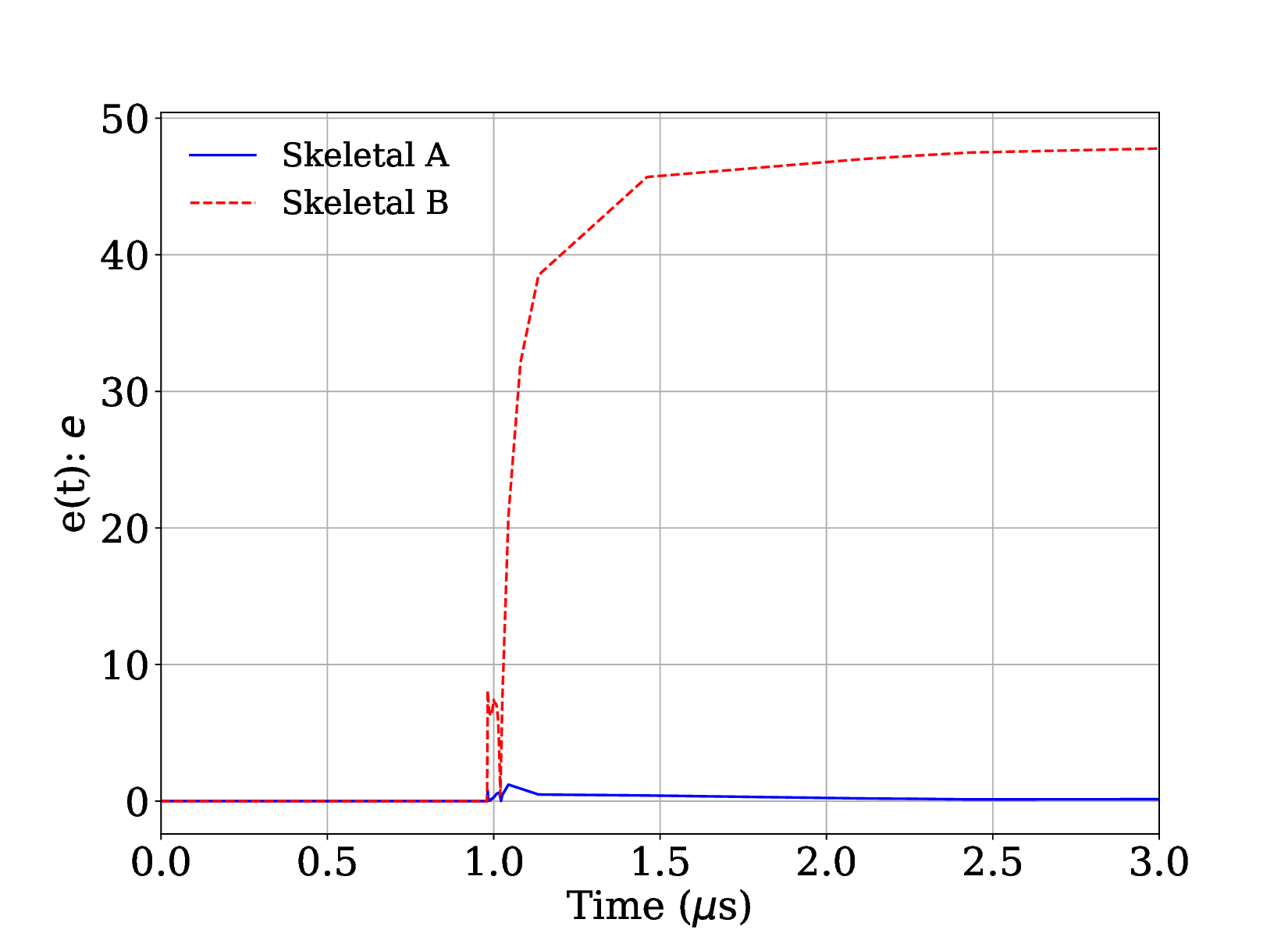}
}
\subfigure[]{
\includegraphics[width=0.50\textwidth, trim=2.0cm 0.0cm 0.0cm 0.0cm]{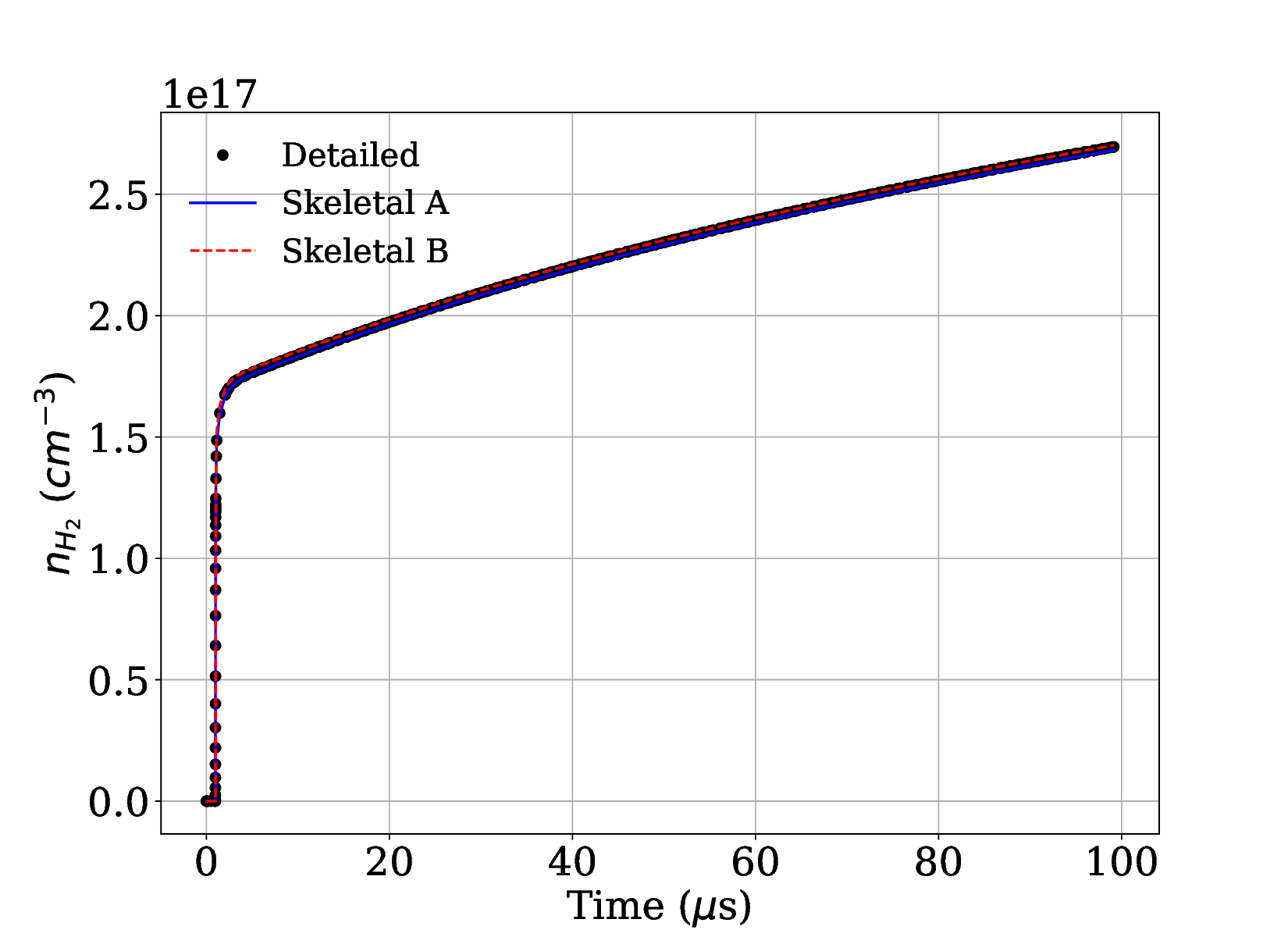}
}
\subfigure[]{
\includegraphics[width=0.50\textwidth, trim=2.0cm 0.0cm 0.0cm 0.0cm]{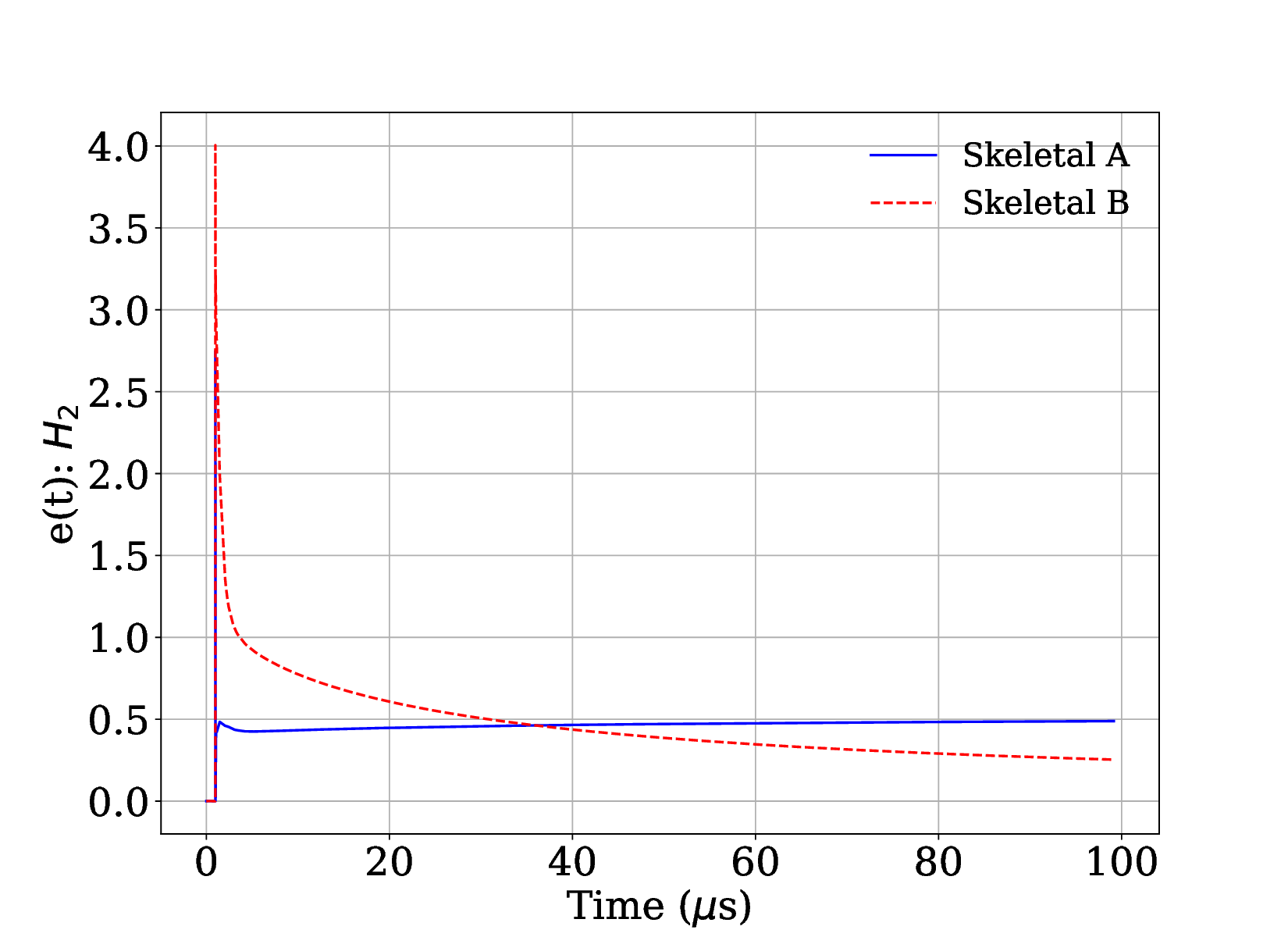}
}
\subfigure[]{
\includegraphics[width=0.50\textwidth, trim=2.0cm 0.0cm 0.0cm 0.0cm]{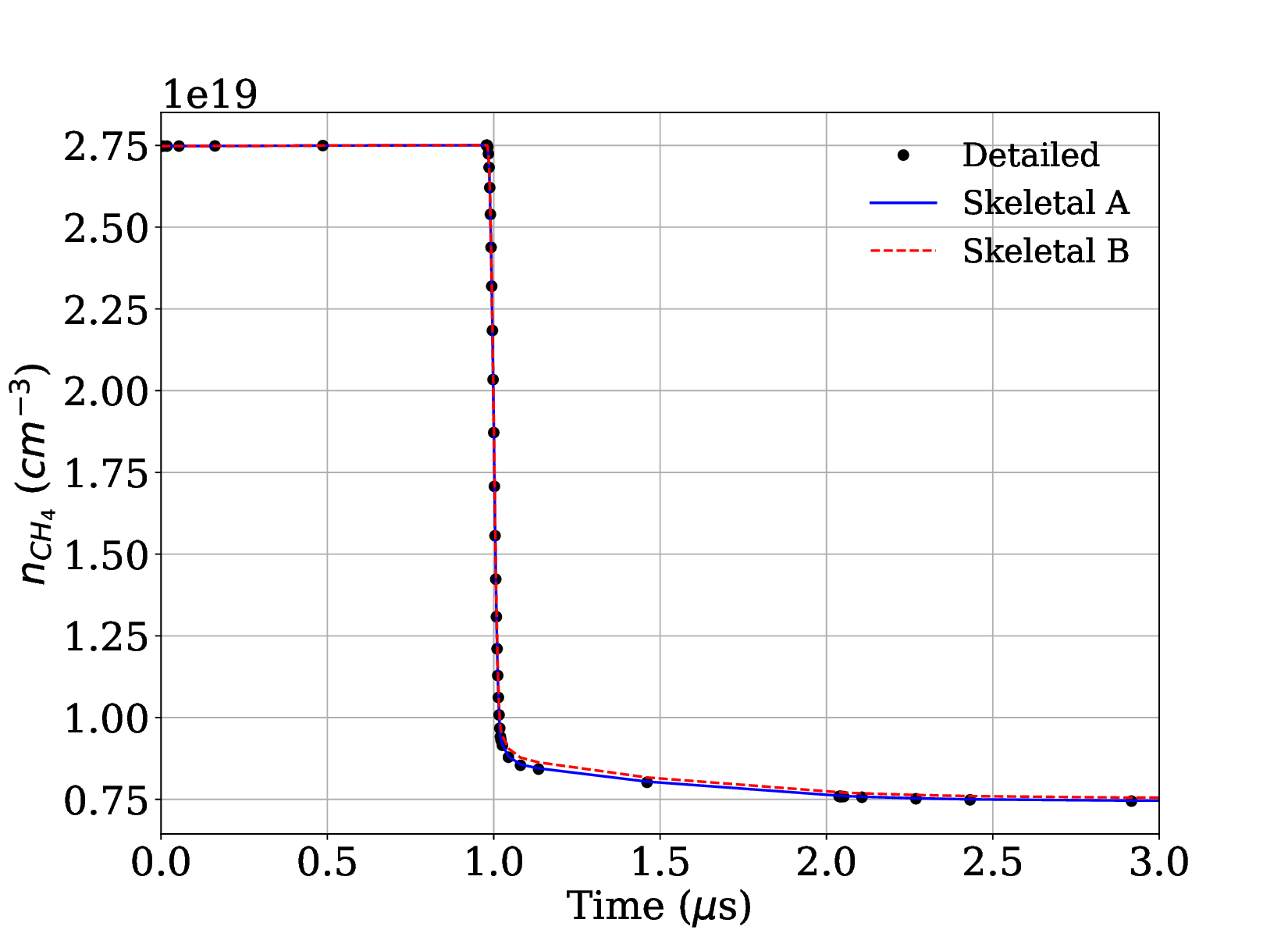}
}
\subfigure[]{
\includegraphics[width=0.50\textwidth, trim=2.0cm 0.0cm 0.0cm 0.0cm]{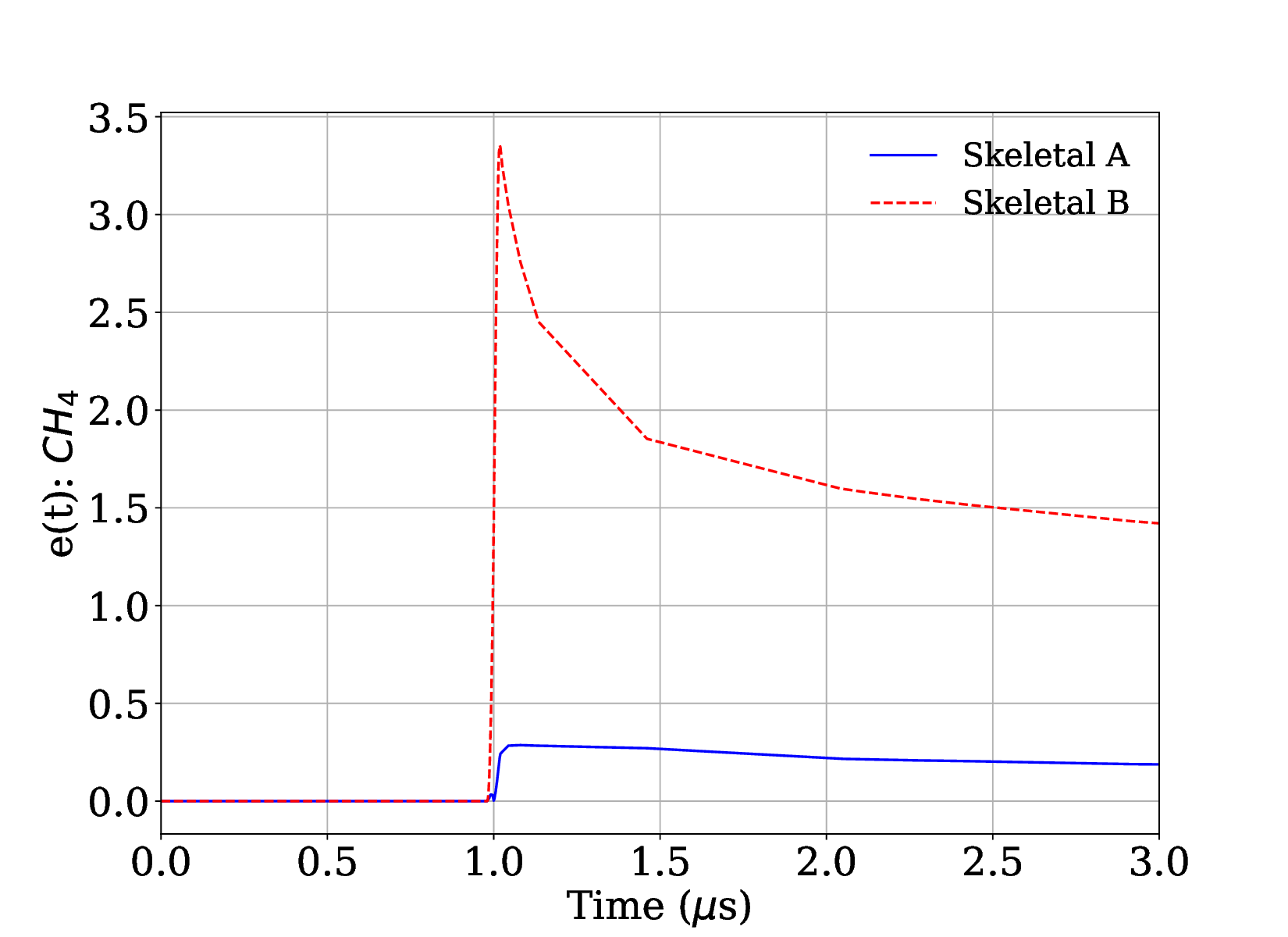}
}
\caption{Number density profiles, and error $e(t)$ (Eq. \ref{eq:localError}) of major target species using the detailed chemical mechanism (77 species), and two skeletal mechanisms: A (32 species), and B (20 species). }
\label{fig:skelVsDetSpeciesA}
\end{figure}

\begin{figure}[h!]
\subfigure[]{
\includegraphics[width=0.50\textwidth, trim=2.0cm 0.0cm 0.0cm 0.0cm]{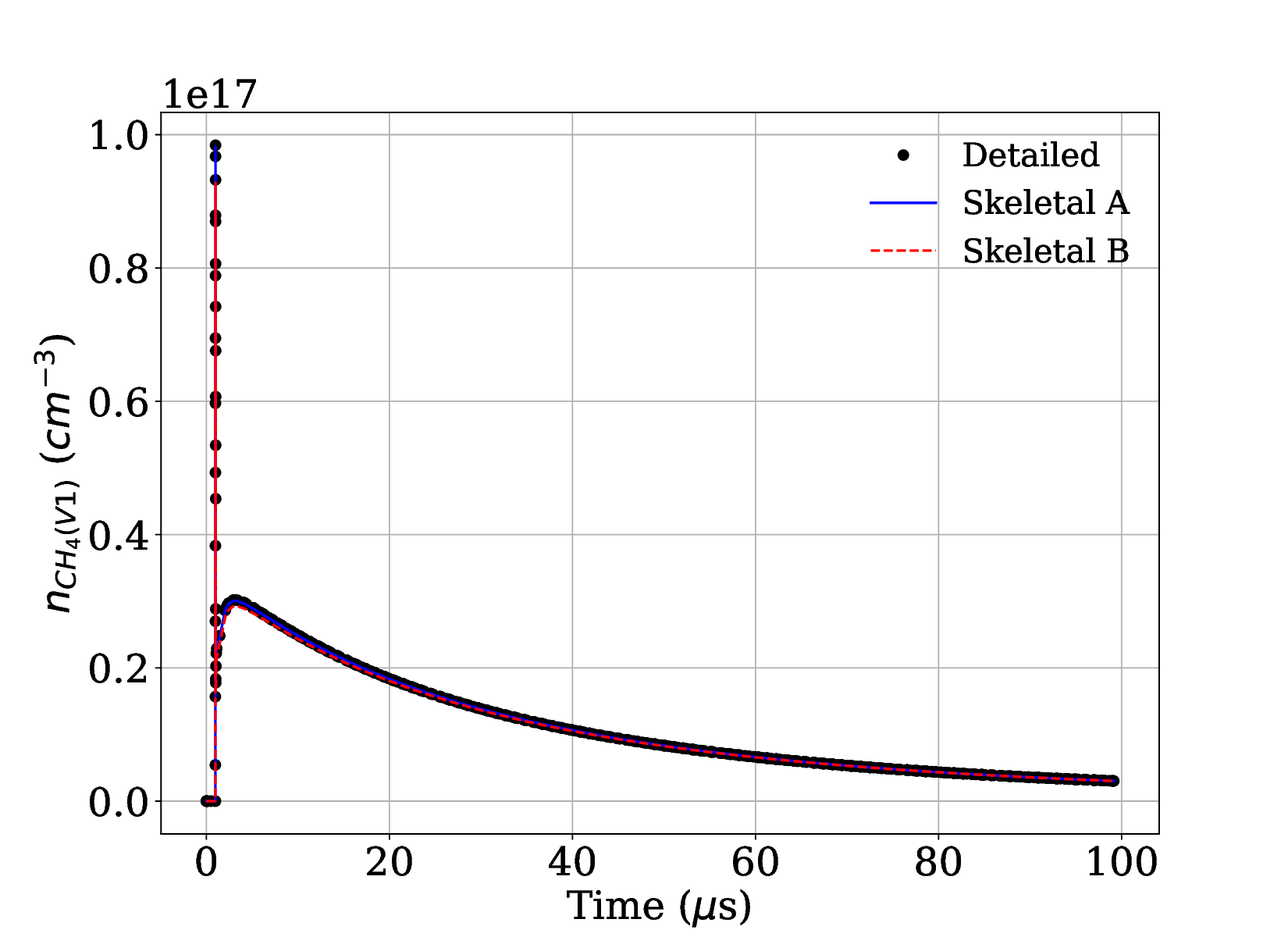}
}
\subfigure[]{
\includegraphics[width=0.50\textwidth, trim=2.0cm 0.0cm 0.0cm 0.0cm]{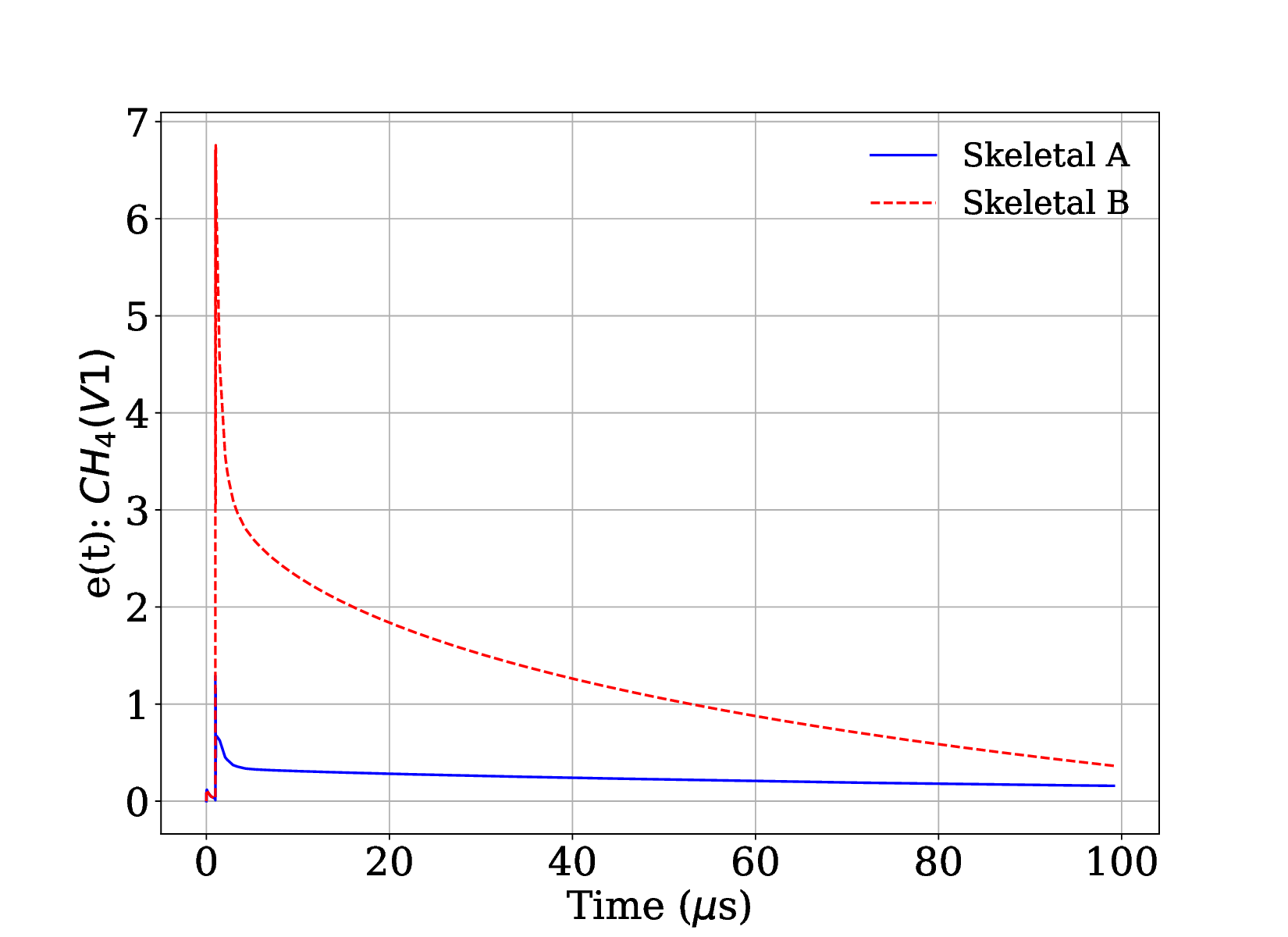}
}
\subfigure[]{
\includegraphics[width=0.50\textwidth, trim=2.0cm 0.0cm 0.0cm 0.0cm]{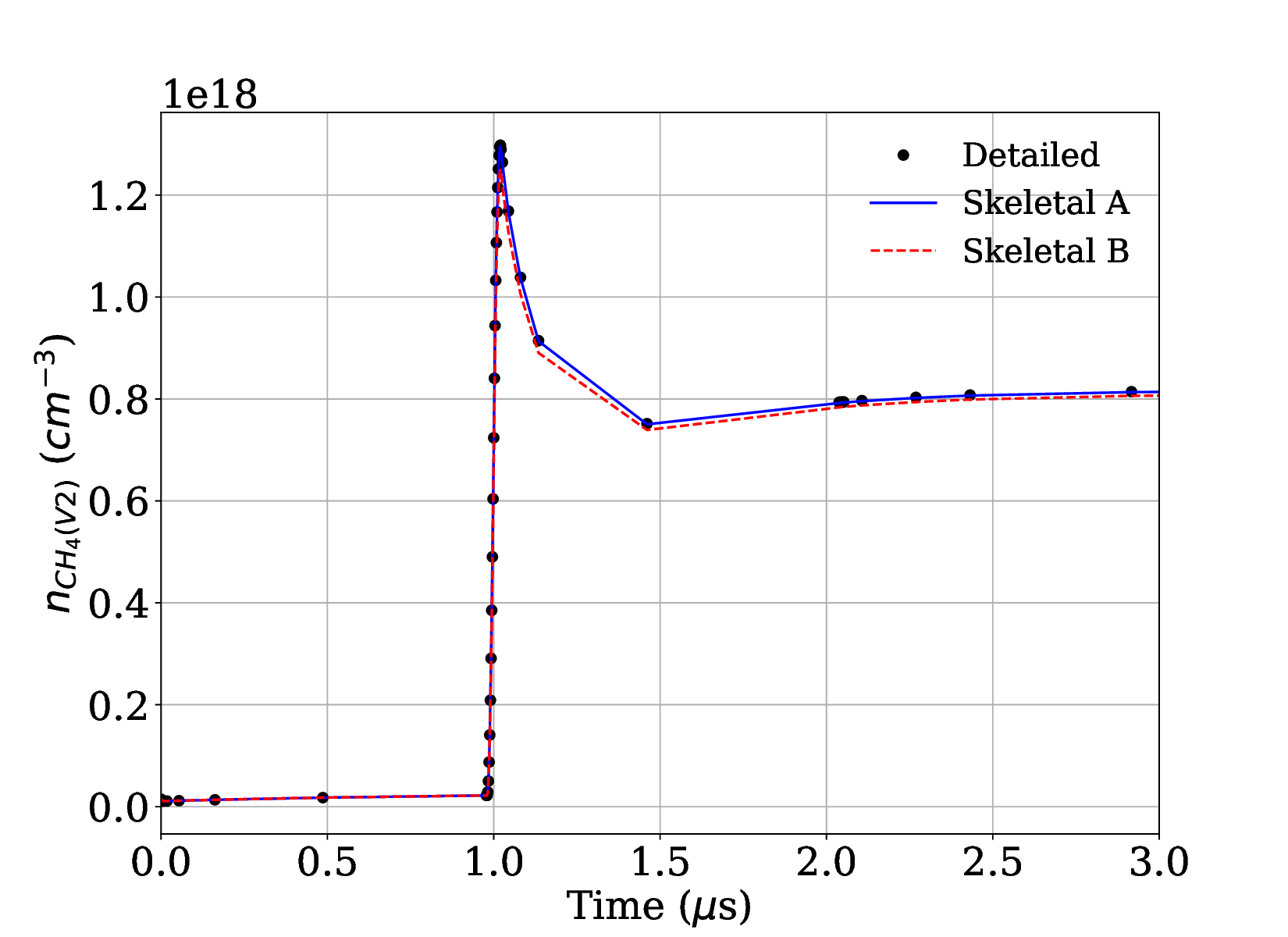}
}
\subfigure[]{
\includegraphics[width=0.50\textwidth, trim=2.0cm 0.0cm 0.0cm 0.0cm]{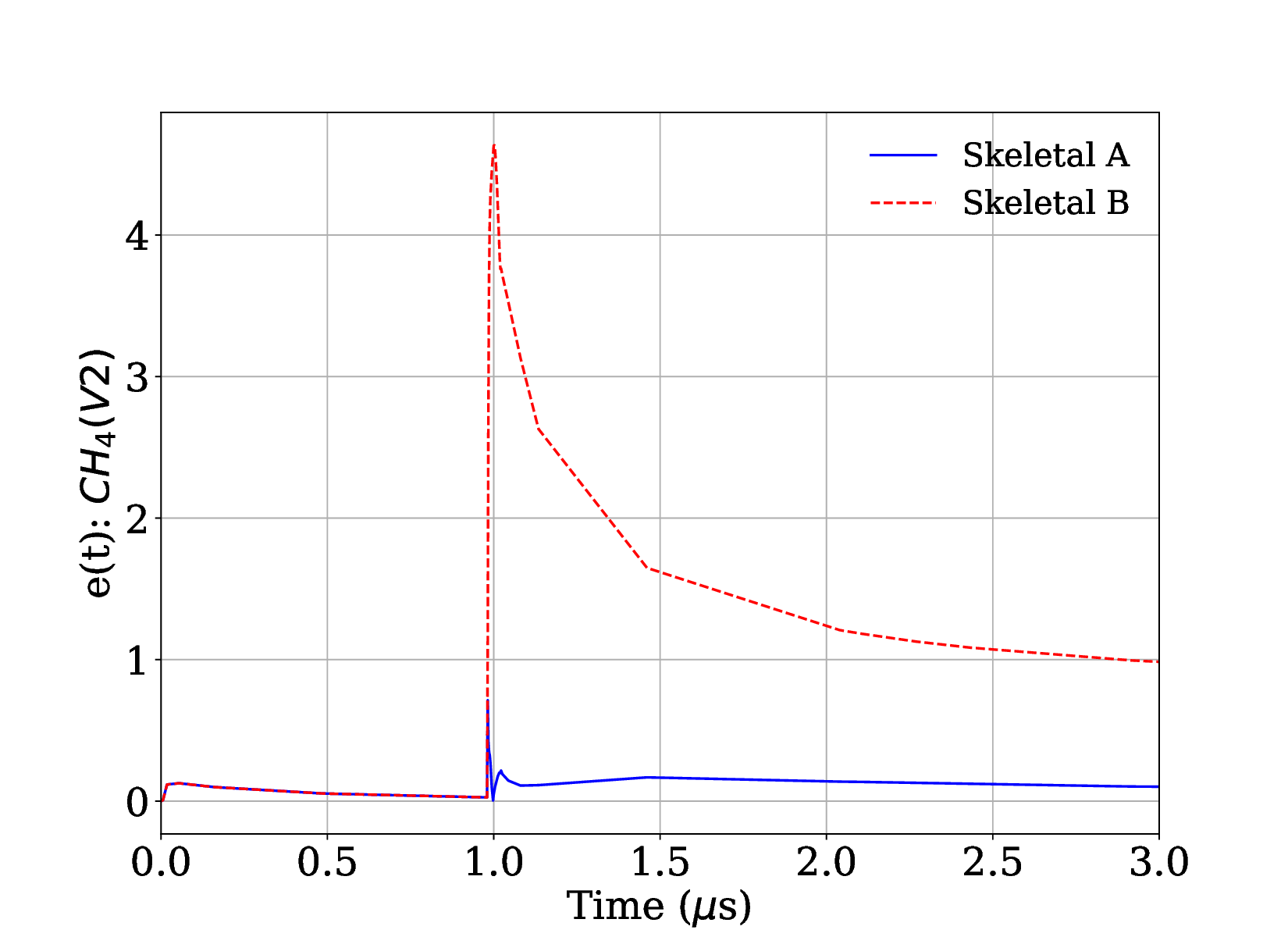}
}
\subfigure[]{
\includegraphics[width=0.50\textwidth, trim=2.0cm 0.0cm 0.0cm 0.0cm]{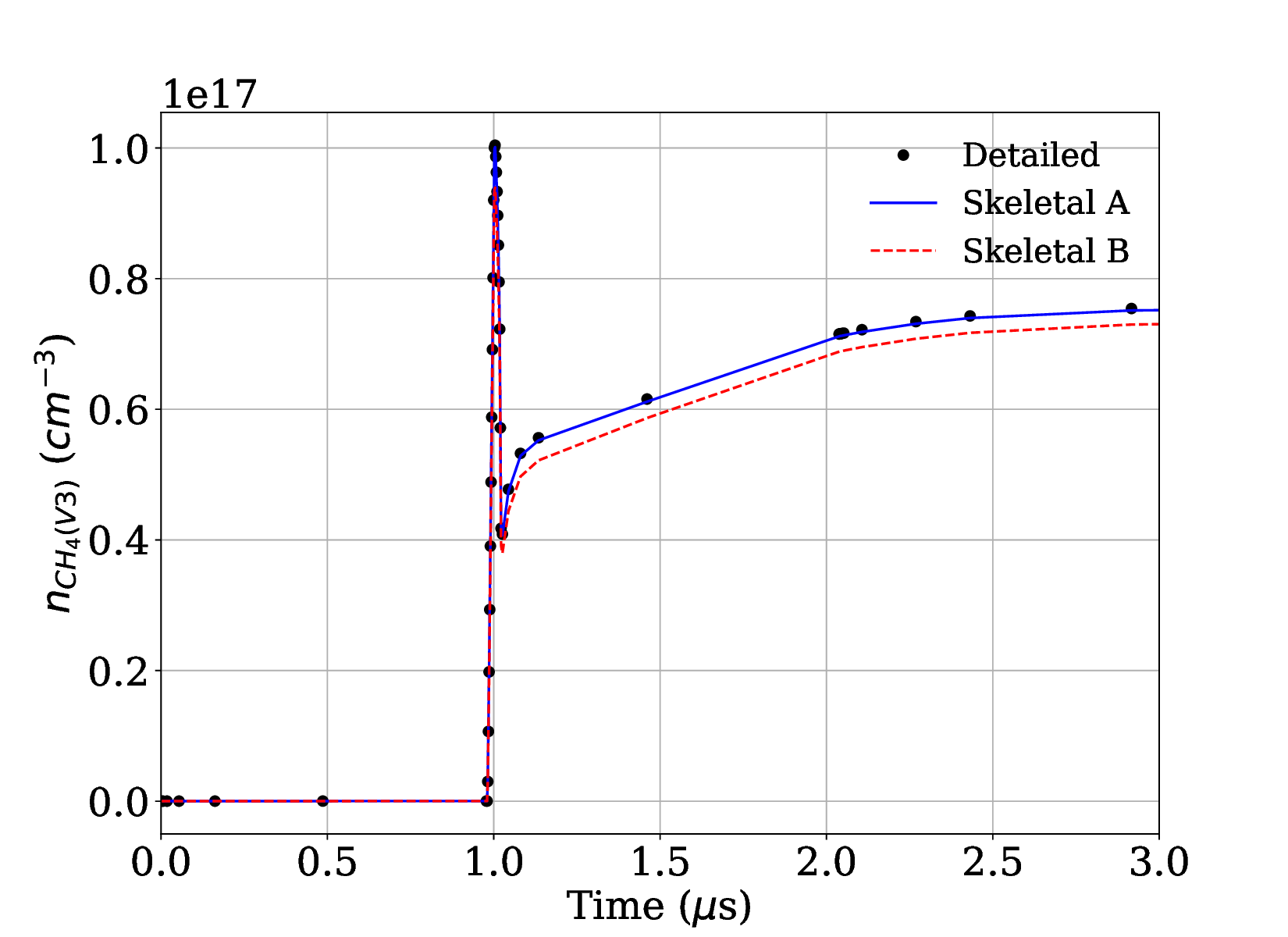}
}
\subfigure[]{
\includegraphics[width=0.50\textwidth, trim=2.0cm 0.0cm 0.0cm 0.0cm]{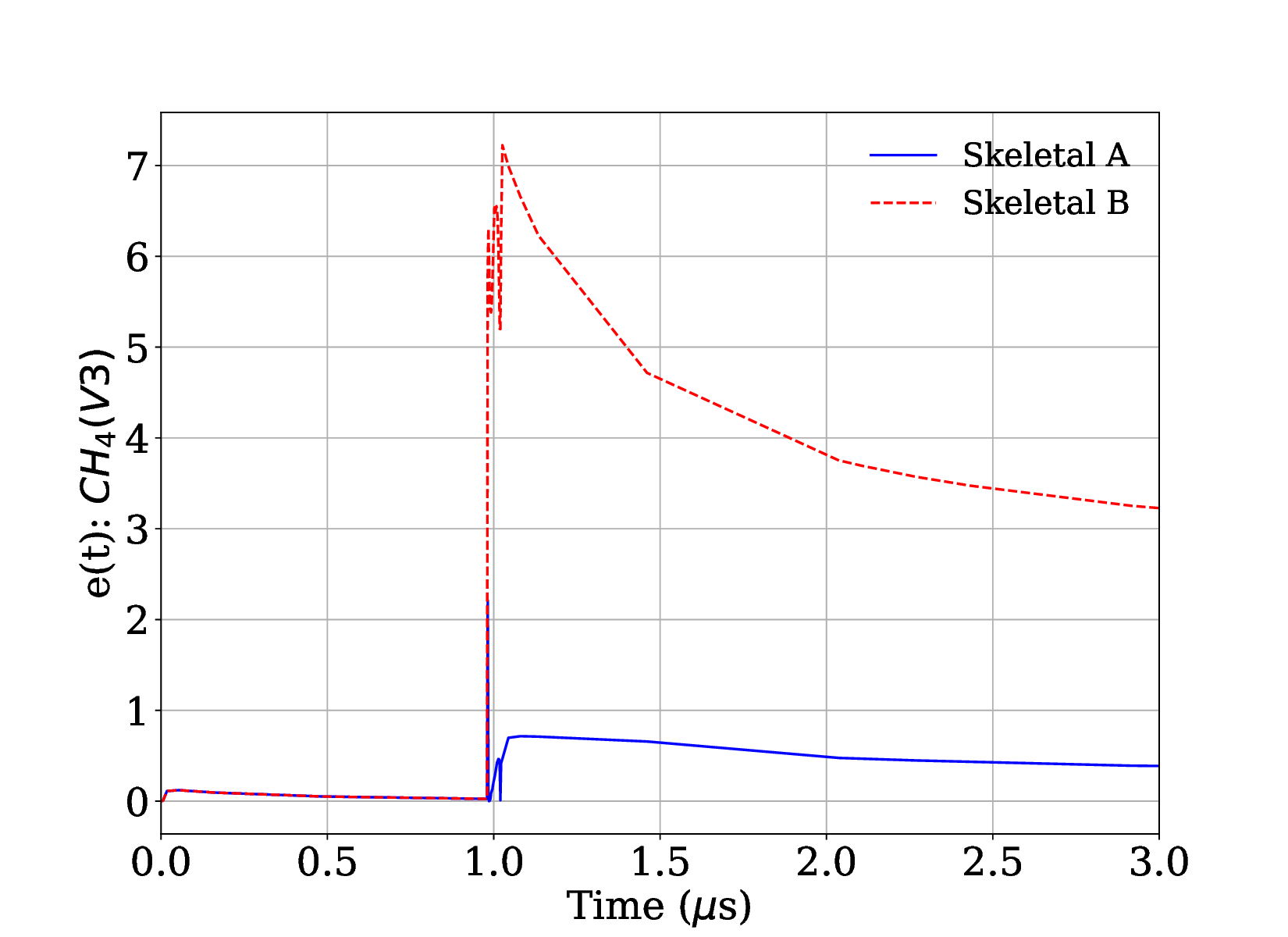}
}
\caption{Number density profiles, and error $e(t)$ (Eq. \ref{eq:localError}) of major target species using the detailed chemical mechanism (77 species), and two skeletal mechanisms: A (32 species), and B (20 species).}
\label{fig:skelVsDetSpeciesB}
\end{figure}

\begin{figure}[h!]
\subfigure[]{
\includegraphics[width=0.50\textwidth, trim=2.0cm 0.0cm 0.0cm 0.0cm]{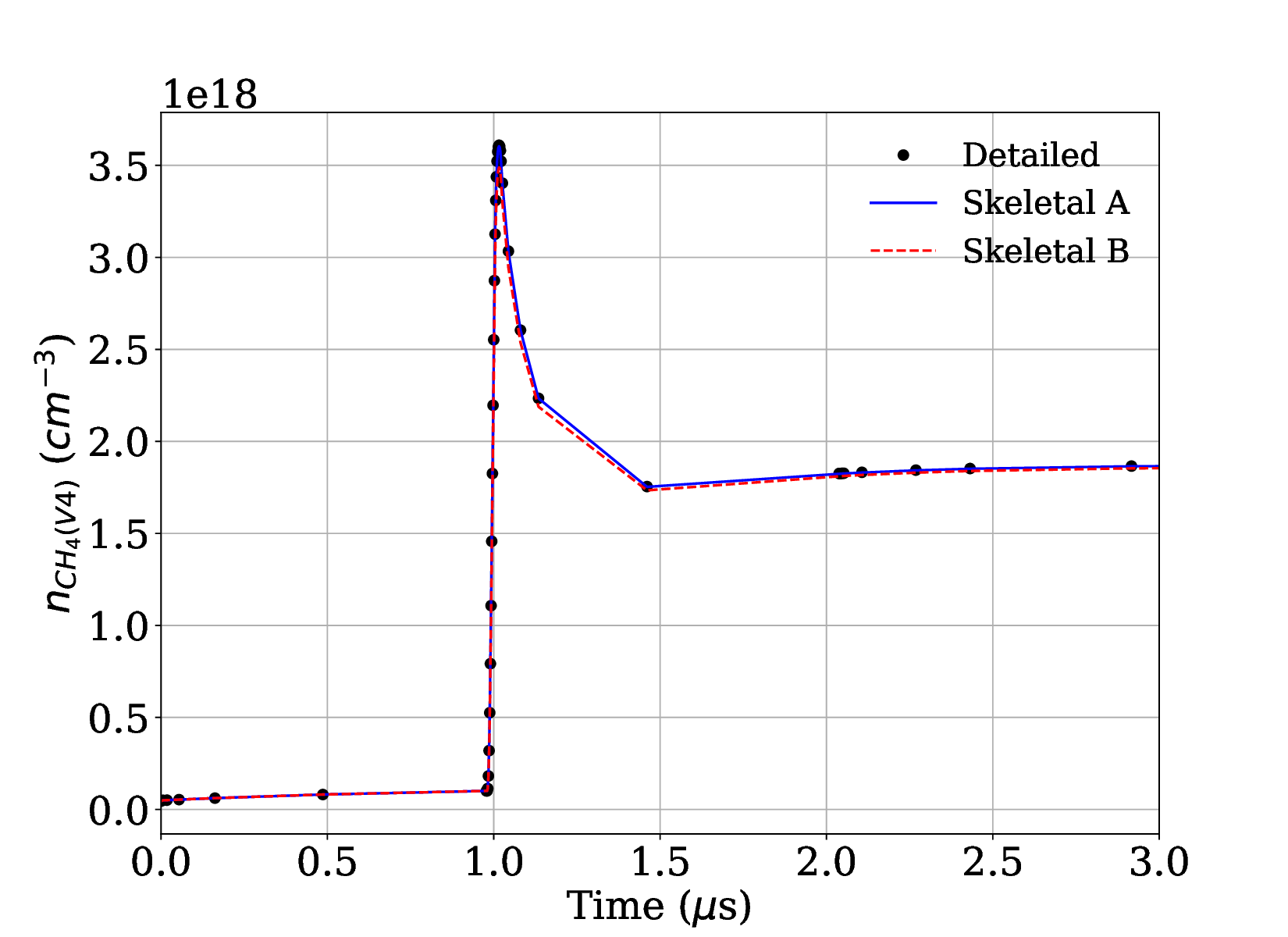}
}
\subfigure[]{
\includegraphics[width=0.50\textwidth, trim=2.0cm 0.0cm 0.0cm 0.0cm]{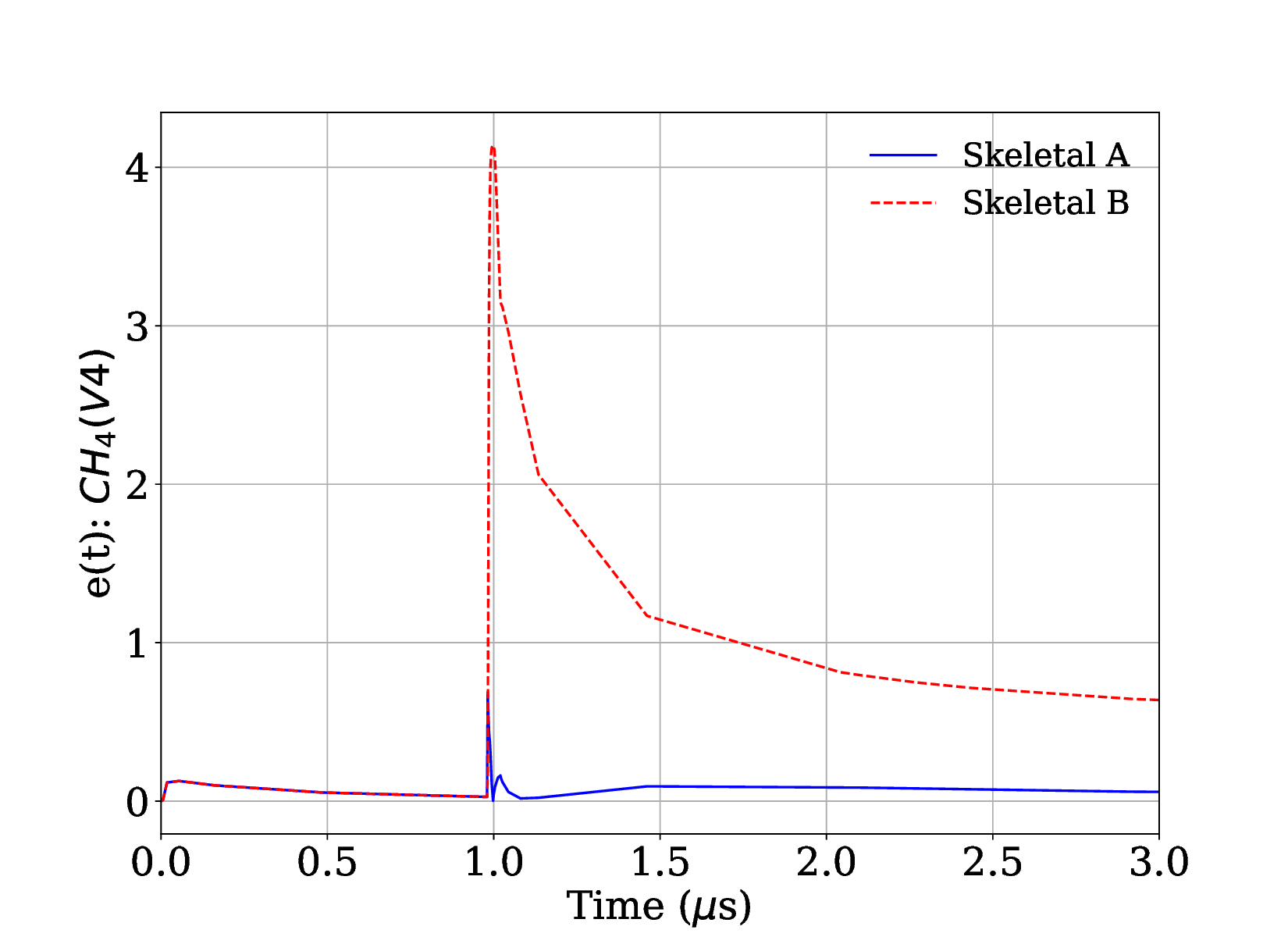}
}
\subfigure[]{
\includegraphics[width=0.50\textwidth, trim=2.0cm 0.0cm 0.0cm 0.0cm]{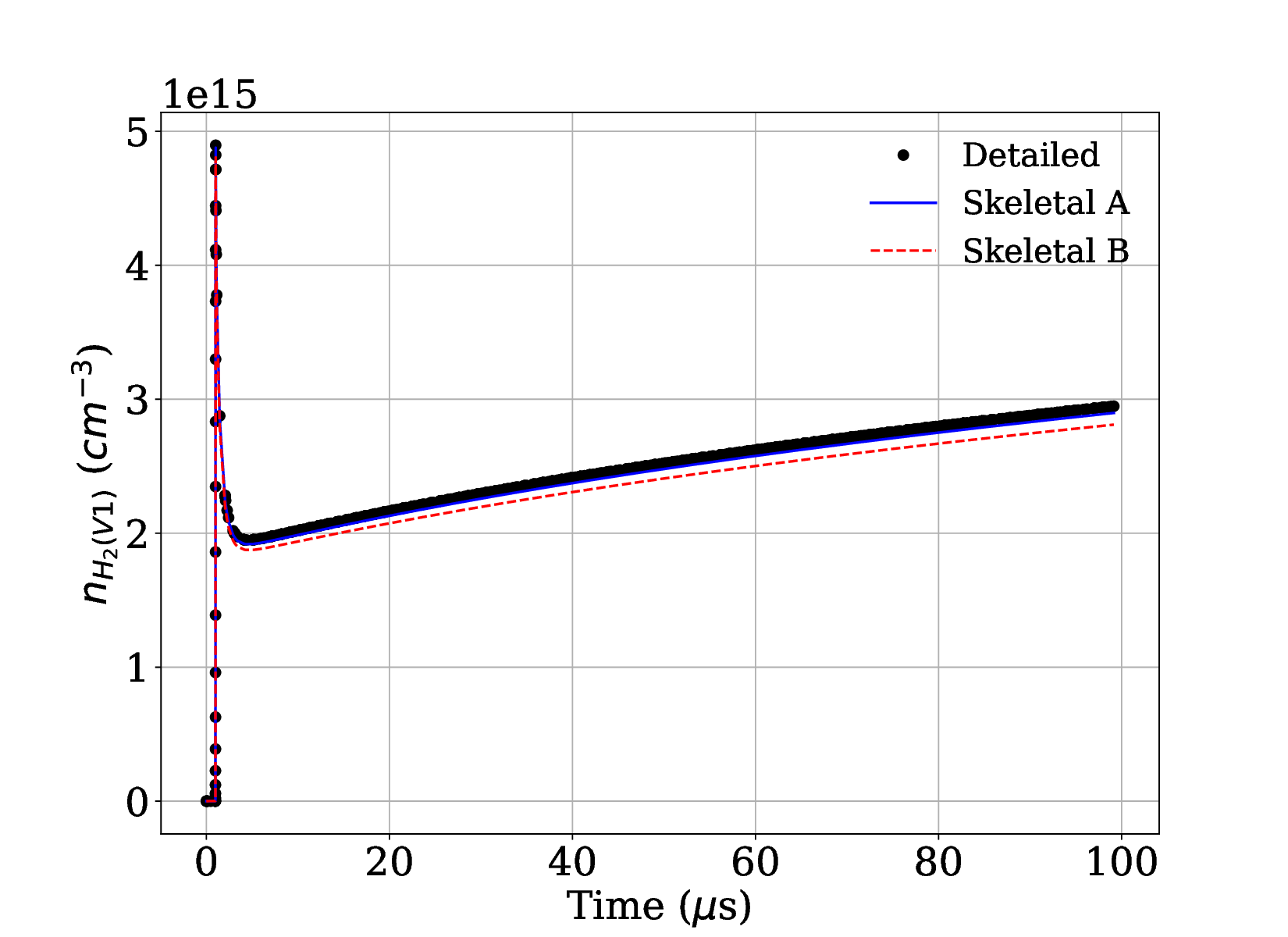}
}
\subfigure[]{
\includegraphics[width=0.50\textwidth, trim=2.0cm 0.0cm 0.0cm 0.0cm]{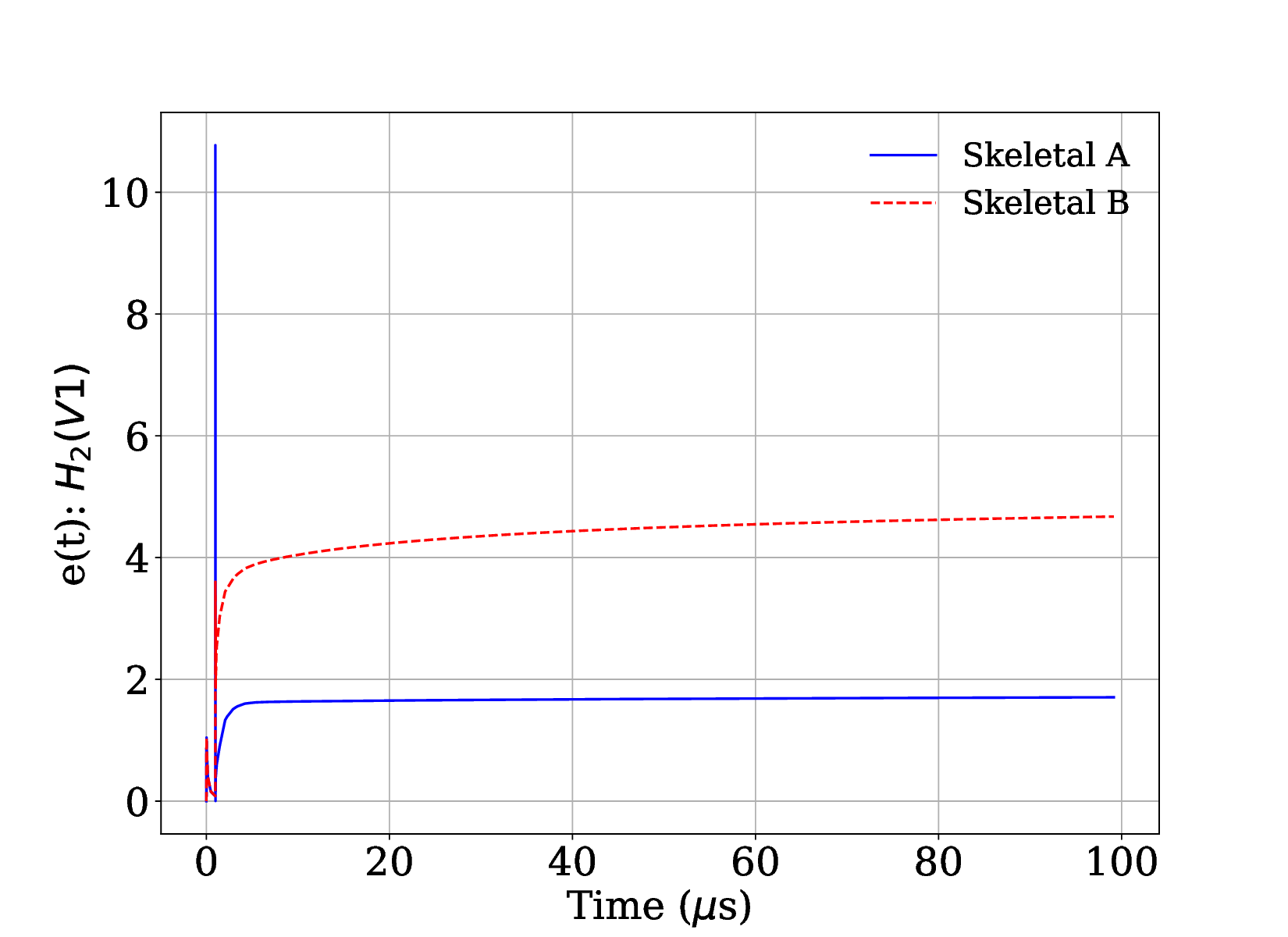}
}
\subfigure[]{
\includegraphics[width=0.50\textwidth, trim=2.0cm 0.0cm 0.0cm 0.0cm]{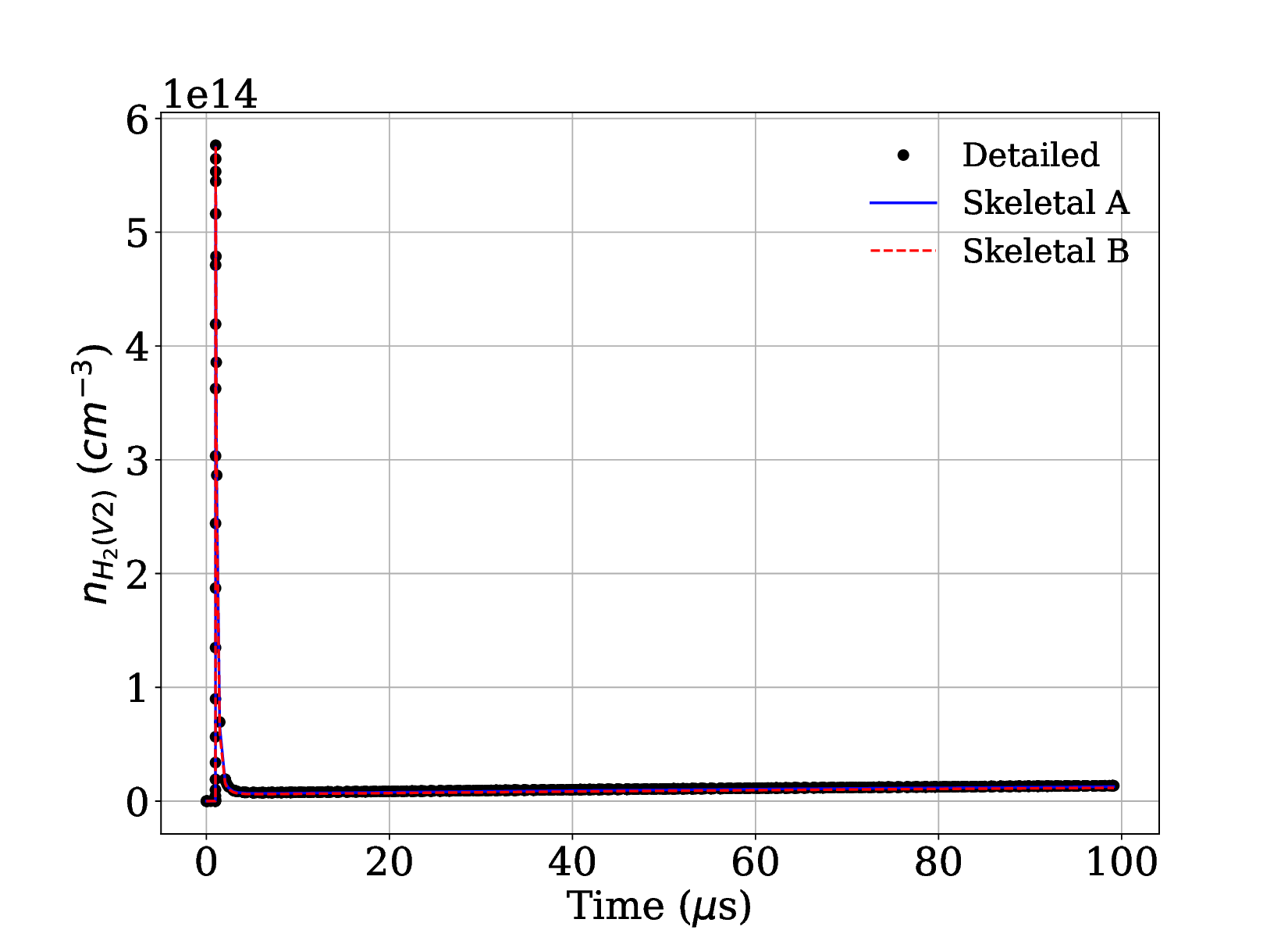}
}
\subfigure[]{
\includegraphics[width=0.50\textwidth, trim=2.0cm 0.0cm 0.0cm 0.0cm]{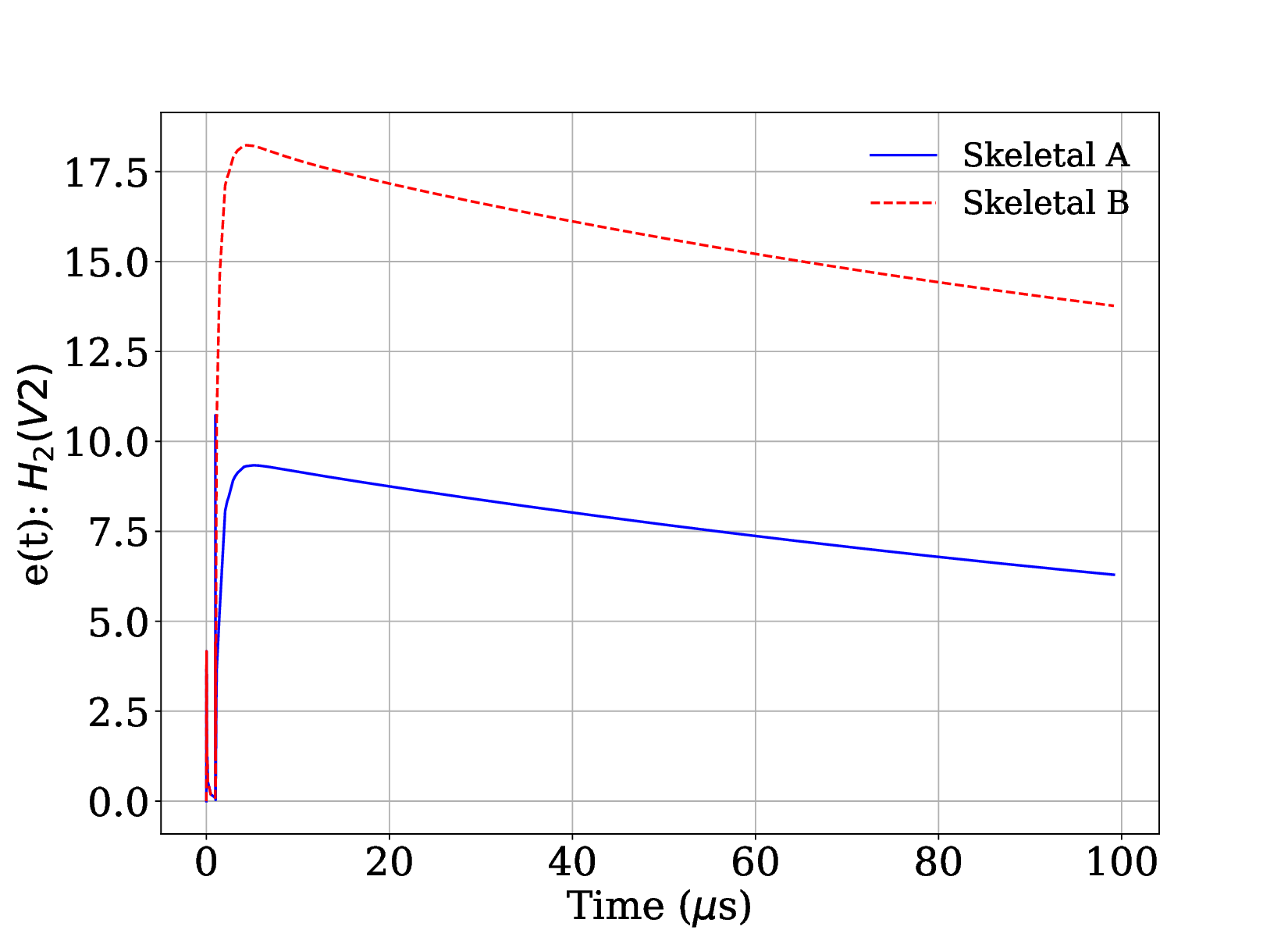}
}
\caption{Number density profiles, and error $e(t)$ (Eq. \ref{eq:localError}) of major target species using the detailed chemical mechanism (77 species), and two skeletal mechanisms: A (32 species), and B (20 species).}
\label{fig:skelVsDetSpeciesC}
\end{figure}

Figure \ref{fig:skelVsDetEandT} shows the corresponding electric field and temperature profiles obtained using the detailed chemistry as well as the two skeletal mechanisms generated using \REDLIB. Clearly, for both skeletal mechanisms the electric field and temperature profiles are in good agreement with the detailed chemistry results despite the large reduction. The corresponding errors are less than 5\% for both, and peak around the point where the electric field and temperature peak. Figures \ref{fig:skelVsDetSpeciesA}-\ref{fig:skelVsDetSpeciesC} further show the number-density profiles as well as the local absolute percentage error, $e(t)$, as defined in Eq. \ref{eq:localError}. For skeletal mechanism A, the local error remains small for all target species, and peaks around the pulse region. For skeletal mechanism B, the error also remains small for all species except for the electrons where the peak error is found to occur only in the afterglow region where the electron density falls sharply to very small values in comparison to the peak value around the pulse region. As the rates of electron-cation recombination reactions in the afterglow period are described by rate expressions, this larger error could be a result of the absence of the $\mathrm{C_2H_3^+}$, $\mathrm{C_2H_4^+}$ and $\mathrm{CH_3^+}$ species (and their individual associated reactions) in skeletal mechanism B, which may lead to the overestimated electron density predictions. Nevertheless, the number-density prediction is still in good agreement with the detailed chemistry results around the pulse region and where the electron density peaks-around this region the error is in fact less than 10\%. It is also important to note at this point that the computational speedup scales with the reduction ratio in the species hence for skeletal mechanism A the speedup factor is 2.0 and for skeletal mechanism B the speedup stands at 3.1. These are both significant speedups while the corresponding loss in accuracy is in comparison of a much lower order. For smaller reduction ratios (more species), the error falls to even lower values as one may see from Fig. \ref{fig:percError}-it is up to the user to select the optimum between reduction level, and accuracy. 

Overall, these results help to validate the implementation of \REDLIB, and at the same time serve to show the potentially large impact the DRGEP method has on reducing large-scale plasma chemical mechanisms and accelerate numerical simulations. It is important to note that in comparison to previous studies in the literature focusing on plasma-chemistry reduction, the chemical mechanism employed in this study is substantially larger and more complex. In addition, DRGEP achieved large reduction factors both in the number of species and in the number of associated reactions while using data obtained from solving a canonical problem at experimentally relevant conditions \cite{2023_chemEngJourn_morais}. Due to its generality, \REDLIB\ may be used to reduce arbitrary large-scale plasma chemical mechanisms given in ZDPlasKin format, and for reaction-rate data obtained for any canonical problem at different applied power-profiles, pressures, temperatures etc. 

\clearpage

\section{Conclusions}

An open-source software library for the automatic reduction of large-scale plasma chemical mechanisms has been developed.   
\REDLIB\ has been validated using data generated by solving a 0D initial-value problem corresponding to a nano-second pulsed power discharge in a homogeneous reactor which generates a methane-hydrogen plasma at experimentally relevant conditions. Gas-heating as well as heat-exchanges with the environment have also been taken into account. The detailed chemical mechanism used is quite large consisting of 77 species and 4404 reactions many of which have complex user-provided reaction rate expressions. Numerous skeletal mechanisms were generated down to 20 species and 442 reactions with substantial computational speedups while achieving a relatively small mean maximum absolute percentage error for the target species number-density profiles. For slightly larger skeletal mechanisms, the error becomes almost insignificant.  

To the best of our knowledge, this is the first open-source, and freely-available automated reduction tool tailored to plasma chemistry, and which at the same time employs a powerful reduction method namely DRGEP \cite{2008_cnf_pepiot}. The library is written in Fortran 90, and has no additional dependencies. \REDLIB\ is meant to be used as a tool by the plasma-physics community for reducing complex detailed plasma chemical mechanisms, accelerating numerical simulations, and gaining understanding into the dominant underlying reaction physics.  

\section*{Authors contribution}

Zacharias Nikolaou: Conceptualization, Methodology, Software, Writing-review \& editing, Writing-original draft, Visualization, Validation, Data curation. Eduardo Morais: Conceptualization, Validation, Writing-review \& editing.
Stijn Van Rompaey: Development of chemical mechanism. Annemie Bogaerts: Conceptualization, Validation, Writing-review \& editing. Charalambos Anastassiou: Funding acquisition, Conceptualization, Writing-review \& editing. Vasileios Vavourakis: Funding acquisition, Conceptualization, Writing-review \& editing.  

\section*{Declaration of competing interest}

The authors declare that they have no known competing financial
interests or personal relationships that could have appeared to influence
the work reported in this paper.

\section*{Acknowledgements}

ZN, CA and VV acknowledge to have received funding by the European Union as part of the European Innovation Council project IgnitePLASMA (Call: HORIZON-EIC-2023-PATHFINDEROPEN-01; grant ID: 101129853).

\bibliographystyle{unsrt}
\bibliography{ref.bib}

\begin{thebibliography}{10}

\bibitem{2008_wiley_goebel}
D.~Goebel and I.~Katz.
\newblock Fundamentals of electric propulsion: {I}on and {H}all {T}hrusters.
\newblock {\em New York: Wiley}, 2008.

\bibitem{1993_applPhysLet_ventzek}
P.L.G. Ventzek, T.J. Sommerer, R.J. Hoekstra, and M.J. Kushner.
\newblock Two‐dimensional hybrid model of inductively coupled plasma sources
  for etching.
\newblock {\em Appl. Phys. Letters}, 63:605–607, 1993.

\bibitem{2016_clinicalPlasmaMed_sander}
S.~Bekeschus, A.~Schmidt, K.D. Weltmann, and T.~von Woedtke.
\newblock The plasma jet k{INP}en–a powerful tool for wound healing.
\newblock {\em Clin. Plasma Med.}, 4:19--28, 2016.

\bibitem{2013_physPlasmas_keidar}
M.~Keidar, A.~Shashurin, O.~Volotskova, M.A. Stepp, P.~Srinivasan, A.~Sandler,
  and B.~Trink.
\newblock Cold atmospheric plasma in cancer therapy.
\newblock {\em Phys. Plasmas}, 20:057101, 2013.

\bibitem{1988_procTech_kogel}
U.~Kogelschatz.
\newblock Advanced ozone generation: {P}rocess {T}echnologies for {W}ater
  {T}reatment.
\newblock {\em Springer US}, pages 87--118, 1988.

\bibitem{2017_chemSoc_snoeckx}
R.~Snoeckx and A.~Bogaerts.
\newblock Plasma technology–a novel solution for {$\mathrm{CO_2}$}
  conversion?
\newblock {\em Chem. Soc. Rev.}, 46:5805–5863, 2017.

\bibitem{2018_acs_bogaerts}
A.~Bogaerts and E.C. Neyts.
\newblock Plasma technology: An emerging technology for energy storage.
\newblock {\em ACS Energy Lett.}, 3:1013--1027, 2018.

\bibitem{2020_heijkers_jPhysChem}
S.~Heijkers, M.~Aghaei, and A.~Bogaerts.
\newblock Plasma-{B}ased {$\mathrm{CH_4}$} {C}onversion into {H}igher
  {H}ydrocarbons and {$\mathrm{H_2}$}: {M}odeling to {R}eveal the {R}eaction
  {M}echanisms of {D}iﬀerent {P}lasma {S}ources.
\newblock {\em J. Phys. Chem.}, 124:7016--7030, 2020.

\bibitem{2021_joule_winter}
L.R. Winter and J.G. Chen.
\newblock {$\mathrm{N_2}$} {F}ixation by {P}lasma-{A}ctivated {P}rocesses.
\newblock {\em Joule}, 5:300--315, 2021.

\bibitem{FULCHERI2024100973}
L.~Fulcheri, E.~Dames, and V.~Rohani.
\newblock Plasma-based conversion of methane into hydrogen and carbon black.
\newblock {\em Current Op. Green and Sust. Chem.}, 50:100973, 2024.

\bibitem{FEDIRCHYK2024155946}
Igor Fedirchyk, Ivan Tsonev, Rubén {Quiroz Marnef}, and Annemie Bogaerts.
\newblock Plasma-assisted {$\mathrm{NH_3}$} cracking in warm plasma reactors
  for green {$\mathrm{H_2}$} production.
\newblock {\em Chem. Eng. J.}, 499:155946, 2024.

\bibitem{2025_natureChem_bogaerts}
A.~Bogaerts.
\newblock Plasma technology for the electrification of chemical reactions.
\newblock {\em Nat. Chem. Eng.}, 2:336–340, 2025.

\bibitem{2022_jPhysD_adamovich}
I.~Adamovich, S.~Agarwal., E~Ahedo, L.L. Alves, S.~Baalrud, N.~Babaeva,
  A.~Bogaerts, A.~Bourdon, P.J. Bruggeman, C.~Canal, E.H. Choi, S.~Coulombe,
  Z.~Donkó, D.B. Graves, S.~Hamaguchi, D.~Hegemann, M.~Hori, H.H Kim, G.M.W.
  Kroesen, M.J. Kushner, A.~Laricchiuta, X.~Li, T.E. Magin, S.~Mededovic
  Thagard, V.~Miller, A.B. Murphy, G.S. Oehrlein, N.~Puac, R.M. Sankaran,
  S.~Samukawa, M.~Shiratani, M.~Simek, N.~Tarasenko, K.~Terashima, E.~Thomas,
  J.~Trieschmann, S.~Tsikata, M.M. Turner, I.J. van~der Walt, M.C.M. van~de
  Sanden, and T.~von Woedtke.
\newblock The 2022 plasma roadmap: low temperature plasma science and
  technology.
\newblock {\em J. Phys. D: Appl. Phys.}, 55:373001, 2022.

\bibitem{2008_plasmaChem_fridman}
A.~Fridman.
\newblock Plasma {C}hemistry.
\newblock {\em Cambridge University Press}, 2008.

\bibitem{2009_jPhysD_kushner}
M.J. Kushner.
\newblock Hybrid modelling of low temperature plasmas for fundamental
  investigations and equipment design.
\newblock {\em J. Phys. D: Appl. Phys.}, 42:194013, 2009.

\bibitem{2018_psst_trelles}
J.P Trelles.
\newblock Advances and challenges in computational fluid dynamics of
  atmospheric pressure plasmas.
\newblock {\em Plasma Sources Sci. Technol.}, 27:093001, 2018.

\bibitem{gaens_jPhysD_2013}
W.~van Gaens and A.~Bogaerts.
\newblock Kinetic modelling for an atmospheric pressure argon plasma jet in
  humid air.
\newblock {\em J. Phys. D: Appl. Phys.}, 46:275201, 2013.

\bibitem{2016_jPhysD_leitz}
A.M. Leitz and M.J. Kushner.
\newblock Air plasma treatment of liquid covered tissue: long timescale
  chemistry.
\newblock {\em J. Phys. D: Appl. Phys.}, 49:425204, 2016.

\bibitem{2018_physChem_schroter}
S.~Schroter, A.~Wijaikhum, A.R. Gibson, A.~West, H.L. Davies, N.~Minesi,
  J.~Dedrick, E.~Wagenaars, N.~de~Oliveira, L.~Nahon, M.J. Kushner, J.P. Booth,
  K.~Niemi, T.~Gans, and D.~O'Connell.
\newblock Chemical kinetics in an atmospheric pressure helium plasma containing
  humidity.
\newblock {\em Phys. Chem. Chem. Phys.}, 20:24263, 2018.

\bibitem{1989_intJChemKin_turanyi}
T.~Turanyi, T.~Berces, and S.~Vajda.
\newblock Reaction rate analysis of complex kinetic systems.
\newblock {\em Int. J. Chem. Kinet.}, 21:83--99, 1989.

\bibitem{2008_ctm_pepiot}
P.~Pepiot-Desjardins and H.~Pitsch.
\newblock An automatic chemical lumping method for the reduction of large
  chemical kinetic mechanisms.
\newblock {\em Combust. Theory Model.}, 12:1089--1108, 2008.

\bibitem{2006_jPhysChem_lu}
T.~Lu and C.K. Law.
\newblock Systematic approach to obtain analytic solutions of {Q}uasi {S}teady
  {S}tate {S}pecies in reduced mechanisms.
\newblock {\em J. Phys. Chem. A}, 110:13202--13208, 2006.

\bibitem{2013_cnf_nikolaou}
Z.~Nikolaou, J.Y. Chen, and N.~Swaminathan.
\newblock A 5-step reduced mechanism for combustion of
  {CO}/{$\mathrm{H_2}$}/{$\mathrm{H_2O}$}/{$\mathrm{CH_4}$}/{$\mathrm{CO_2}$}
  mixtures with low hydrogen/methane and high {$\mathrm{H_2O}$} content.
\newblock {\em Combust. Flame}, 160:56--75, 2013.

\bibitem{2014_cnf_nikolaou}
Z.~Nikolaou, N.~Swaminathan, and J.Y. Chen.
\newblock Evaluation of a reduced mechanism for turbulent premixed combustion.
\newblock {\em Combust. Flame}, 161:3085--3099, 2014.

\bibitem{2005_procCombustInst_lu}
T.~Lu and C.K. Law.
\newblock A directed relation graph method for mechanism reduction.
\newblock {\em Proc. Combust. Inst.}, 30:1333--1341, 2005.

\bibitem{2008_cnf_pepiot}
P.Pepiot Desjardins and H.~Pitsch.
\newblock An efficient error-propagation-based reduction method for large
  chemical kinetic mechanisms.
\newblock {\em Combust. Flame}, 154:67--81, 2008.

\bibitem{1992_cnf_maas}
U.~Maas and S.B. Pope.
\newblock Simplifying chemical kinetics: {I}ntrinsic low-dimensional manifolds
  in composition space.
\newblock {\em Combust. Flame}, 88:239--264, 1992.

\bibitem{1994_intJChemKin_lam}
S.H. Lam and D.A. Goussis.
\newblock The {CSP} method for simplifying kinetics.
\newblock {\em Int. J. Chem. Kinet.}, 26:461--486, 1994.

\bibitem{2009_pca_cnf_sutherland}
J.C. Sutherland and A.~Parente.
\newblock Combustion modeling using principal component analysis.
\newblock {\em Combust. Flame}, 32:1563--1570, 2009.

\bibitem{2020_anns_cnf_wan}
K.~Wan, C.~Barnaud, L.~Vervisch, and P.~Domingo.
\newblock Chemistry reduction using machine learning trained from non-premixed
  micro-mixing modeling: Application to {DNS} of a syngas turbulent oxy-flame
  with side-wall effects.
\newblock {\em Combust. Flame}, 220:119--129, 2020.

\bibitem{furst_compPhysComm_2021}
M.~Furst, A.~Bertolino, A.~Cuoci, T.~Faravelli, A.~Frassoltadi, and A.~Parente.
\newblock Optismoke++: A toolbox for optimization of chemical kinetic
  mechanisms.
\newblock {\em Comput. Phys. Commun.}, 264:107940, 2021.

\bibitem{niemeyer_compPhysComm_2017}
K.E. Niemeyer, N.~J. Curtis, and C.J. Sung.
\newblock pyjac: Analytical jacobian generator for chemical kinetics.
\newblock {\em Comput. Phys. Commun.}, 215:188--203, 2017.

\bibitem{curts_compPhysComm_2022}
N.J. Curtis, K.E. Niemeyer, and C.J. Sung.
\newblock Accelerating reactive-flow simulations using vectorized chemistry
  integration.
\newblock {\em Comput. Phys. Commun.}, 278:108409, 2022.

\bibitem{rao_compPhysComm_2024}
S.~Rao, B.~Chen, and X.~Xu.
\newblock Heterogeneous {CPU}-{GPU} parallelization for modeling supersonic
  reacting flows with detailed chemical kinetics.
\newblock {\em Comput. Phys. Commun.}, 300:109188, 2024.

\bibitem{danciu_compPhysComm_2025}
B.A. Danciu and C.E. Frouzakis.
\newblock Kineti{X}: A performance portable code generator for chemical
  kinetics and transport properties.
\newblock {\em Comput. Phys. Commun.}, 310:109504, 2025.

\bibitem{2010_ctm_pope_isat}
S.B. Pope.
\newblock Computationally efficient implementation of combustion chemistry
  using in situ adaptive tabulation.
\newblock {\em Combust. Theory Model.}, pages 41--63, 2010.

\bibitem{dechant_compPhysComm_2023}
C.~DeChant, C.~Icenhour, S.~Keniley, G.~Gall, A.~Lindsay, D.~Curreli, and
  S.~Shannon.
\newblock Verification and validation of the open-source plasma fluid code:
  Zapdos.
\newblock {\em Comput. Phys. Commun.}, 291:108837, 2023.

\bibitem{verma_compPhysComm_2021}
A.K Verma and A.~Venkattraman.
\newblock {SOMAFOAM}: An {O}pen{F}oam based solver for continuum simulations of
  low-temperature plasmas.
\newblock {\em Comput. Phys. Commun.}, 263:107855, 2021.

\bibitem{zdplaskin}
S.~Pancheshnyi, B.~Eismann, G.J.M. Hagelaar, and L.C. Pitchford.
\newblock Computer code {ZDP}las{K}in:
  \url{http://www.zdplaskin.laplace.univ-tlse.fr}.
\newblock {\em University of Toulouse, LAPLACE, CNRS-UPS-INP, Toulouse,
  France}, 2008.

\bibitem{pinhao_compPhysComm_2001}
N.R. Pinhao.
\newblock {PLASMAKIN}: A chemical kinetics library for plasma physics modeling.
\newblock {\em Comput. Phys. Commun.}, 135:105--131, 2001.

\bibitem{bolsigSolver}
G.J.M. Hagelaar and L.C. Pitchford.
\newblock Solving the {B}oltzmann equation to obtain electron transport
  coefficients and rate coefficients for fluid models.
\newblock {\em Plasma Sources Sci. Technol.}, 14:722--733, 2005.

\bibitem{2004_jAtmosphChem_lehman}
R.~Lehmann.
\newblock An algorithm for the determination of all significant pathways in
  chemical reaction systems.
\newblock {\em J. Atmosph. Chem.}, 47,:45–78, 2004.

\bibitem{markosyan_compPhysComm_2014}
A.H. Markosyan, A.~Luque, F.J. Gordillo-Vazquez, and U.~Ebert.
\newblock Pump{K}in: A tool to find principal pathways in plasma chemical
  models.
\newblock {\em Comput. Phys. Commun.}, 185:2697--2702, 2014.

\bibitem{2022_jPhysD_mousavi}
S.T. Mousavi, J.G.M. Gulpen, W.A.A.D.~Graef amd P.M.J.~Koelman, E.A.D. Carbone,
  and J.~van Dijk.
\newblock Assessment of the suitability of the chemical reaction pathway
  algorithm as a reduction method for plasma chemistry.
\newblock {\em J. Phys. D: Appl. Phys.}, 55:505201, 2022.

\bibitem{2015_pcaPlasma_psst_peerenboom}
K.~Peerenboom, A.~Parente, Tomas Kozak, A.~Bogaerts, and G.~Degrez.
\newblock Dimension reduction of non-equilibrium plasma kinetic models using
  principal component analysis.
\newblock {\em Plasma Sources Sci. Tehcnol.}, 24:025004, 2015.

\bibitem{2016_jPhysConf_rehman}
T.~Rehman, E.~Kemaneci, W.~Graef, and J.~van Dijk.
\newblock Simplifying plasma chemistry via {ILDM}.
\newblock {\em J. Phys.: Conf. Ser.}, 682:012035, 2016.

\bibitem{2020_psst_sun}
S.R. Sun, H.X. Wang, and A.~Bogaerts.
\newblock Chemistry reduction of complex {$\mathrm{CO_2}$} chemical kinetics:
  Application to a gliding arc plasma.
\newblock {\em Plasma Sources Sci. Technol.}, 29:025012, 2020.

\bibitem{MAERIVOET2024152006}
S.~Maerivoet, I.~Tsonev, J.~Slaets, F.~Reniers, and A.~Bogaerts.
\newblock Coupled multi-dimensional modelling of warm plasmas: Application and
  validation for an atmospheric pressure glow discharge in
  {$\mathrm{CO_2}$}/{$\mathrm{CH_4}$}/{$\mathrm{O_2}$}.
\newblock {\em Chem. Eng. J.}, 492:152006, 2024.

\bibitem{nikolaou_gmd_2018}
Z.~Nikolaou, J.Y. Chen, Y.~Proestos, J.~Lelieveld, and R.~Sander.
\newblock Accelerating simulations using ${REDCHEM}_v{0.0}$ for atmospheric
  chemistry mechanism reduction.
\newblock {\em Geosc. Model Devel.}, 11:3391–3407, 2018.

\bibitem{2016_cnf_chen}
Y.~Chen and J.H. Chen.
\newblock Application of {J}acobian defined direct interaction coefficient in
  {DRGEP}-based chemical mechanism reduction methods using different graph
  search algorithms.
\newblock {\em Combust. Flame}, 174:77--84, 2016.

\bibitem{1959_numerMath_Dijkstra}
E.W. Dijkstra.
\newblock A note on two problems in connexion with graphs.
\newblock {\em Numer. Math.}, 1:269–271, 1959.

\bibitem{2023_chemEngJourn_morais}
E.~Morais, E.~Delikonstantis, M.~Scapinello, G.~Smith, G.D. Stefanidis, and
  A.~Bogaerts.
\newblock Methane coupling in nanosecond pulsed plasmas: Correlation between
  temperature and pressure and effects on product selectivity.
\newblock {\em Chem. Eng. Journ.}, 462:142227, 2023.

\bibitem{2024_ppp_morais}
E.~Morais and A.~Bogaerts.
\newblock Modelling the dynamics of hydrogen synthesis from methane in
  nanosecond-pulsed plasmas.
\newblock {\em Plasma Process. Polym.}, 21:e2300149, 2024.

\end{thebibliography}

\end{document}